\RequirePackage{fix-cm}

\documentclass[smallcondensed]{svjour3}

\smartqed

\usepackage{graphicx}
\usepackage{natbib}

\usepackage[LGR,T1]{fontenc}

\usepackage{amsmath}
\usepackage{amssymb}
\usepackage{mathrsfs}
\usepackage{textcomp}

\usepackage{tikz}
\usepackage{float}

\usepackage{geometry}

\usepackage{subfig}

\usepackage{color}

\usepackage{hyperref}
\hypersetup{colorlinks=true,citecolor=blue,linkcolor=blue}

\usepackage{enumerate}

\newcommand{\Ham}{\mathcal{H}}

\newcommand{\fat}[1]{$\displaystyle{#1}$}
\newcommand{\drond}[2]{\frac{\partial #1}{\partial #2}}
\newcommand{\vect}[1]{\mathbf{\boldsymbol{#1}}}
\newcommand{\accolade}{\phantom{\frac{1}{1}\!\!\!\!\!}}

\newcommand{\R}{\mathcal{R}}
\newcommand{\place}{\;\;\;\;\;}

\newcommand{\LGR}{\fontencoding{LGR}\selectfont}
\newcommand{\Latin}{\fontencoding{\encodingdefault}\selectfont}
\newcommand{\sampi}{\text{\LGR \textsampi{} \Latin}\!\!}

\newcommand{\qoppa}{\text{\LGR \textqoppa{} \Latin}\!\!}

\newcommand{\be}{\begin{equation}}
\newcommand{\ee}{\end{equation}}
\newcommand{\ab}{\bar{a}}
\newcommand{\gO}{\mathcal{O}}
\newcommand{\eps}{\varepsilon}
\newcommand{\Xb}{\bar{X}}

\graphicspath{{./}}
\newcommand{\bibpath}{./}

\title{\textbf{An analytical model for tidal evolution in co-orbital systems}}
\subtitle{I. Application to exoplanets}
\author{J\'er\'emy Couturier \and Philippe Robutel \and Alexandre C. M. Correia}
\institute{J. Couturier, P. Robutel \at
	IMCCE, UMR8028 CNRS, Observatoire de Paris, PSL Univ., Sorbonne Univ., 77 av. Denfert-Rochereau, 75014 Paris, France\\
	\email{jeremy.couturier@obspm.fr, philippe.robutel@obspm.fr}           
	\and
	A. C. M. Correia \at
	CFisUC, Departamento de F\'isica, Universidade de Coimbra, 3004-516 Coimbra, Portugal\\
	\email{alexandre.correia@uc.pt}
}
\date{Received: March 25, 2021 / Accepted: May 30, 2021}

\begin{document}
	
\maketitle

\setcounter{tocdepth}{3}


\begin{abstract}
Close-in co-orbital planets (in a 1:1 mean motion resonance) can experience strong tidal interactions with the central star.
Here, we develop an analytical model adapted to the study of the tidal evolution of those systems.
We use a Hamiltonian version of the constant time-lag tidal model, which extends the Hamiltonian formalism developed for the point-mass case. 
We show that co-orbital systems undergoing tidal dissipation either favour the Lagrange or the anti-Lagrange configurations, depending on the system parameters. 
However, for all range of parameters and initial conditions, both configurations become unstable, although 
the timescale for the destruction of the system can be 
larger than the lifetime of the star.
We provide an easy-to-use criterion to determine if an already known close-in exoplanet may have an undetected co-orbital companion.
\end{abstract}

\section{Introduction}
\label{intro}

In the framework of the three-body problem, planets in the co-orbital resonance correspond to a system in a 1:1 mean motion resonance. In other words, the planets have the same mean orbital period, which means that the difference of their mean longitudes librates, generally around a value close to $\pm60^\circ$ or equal to $180^\circ$.  Beyond the famous collinear and equilateral configurations described respectively by \cite{Euler1764} and \cite{Lagrange1772}, other types of orbits are possible in this resonance, making the co-orbital dynamics very rich. While the Euler configurations, denoted by $L_1$, $L_2$ and $L_3$ in the restricted three-body problem, are unstable for all mass ranges, the Lagrange equilibria are linearly stable, provided that the three masses satisfy the relation $27(m_0m_1 + m_0m_2 + m_1m_2) < (m_0 + m_1 + m_2)^2 $ established by \cite{Ga1843} for circular orbits. More general criteria exist for equilateral eccentric configurations \cite[see][]{Dan1964,Robe2002,Nauenberg2002}.  
For the previous inequality to be fulfilled, it is necessary that one of the masses is dominating.  Thus, we denote by $m_0$  the mass of the star,  which is much larger than that of the planets $m_1$ and $m_2$. With these notations, the Gascheau condition is satisfied when  $(m_1 + m_2)/m_0\lesssim1/27$.
Therefore, in the planar case and for small eccentricities, when the sum of the planetary masses is smaller than about \fat{1/27} of the stellar mass, tadpole orbits arise, allowing the difference in the mean longitude to librate around $\pm60^\circ$, with a maximum amplitude that increases as the sum of the planetary masses decreases.
When $(m_1+m_2)/m_0\lesssim3\times10^{-4}$, horseshoe-shaped orbits can arise \cite[see][]{LauCha2002}. They librate around $180^\circ$ with a very large  amplitude of at least $312^\circ$.
For moderate to large eccentricities, quasi-satellite orbits are also possible, for which the planets appear to revolve around each other \cite[see][]{Guippone2010,PousseRobutelVienne2017}, while many other exotic trajectories exist at high eccentricity \citep{LeRoCo2018}.
The dynamics of the inclined problem is even more complex by allowing, among other things, transitions between the different types of orbits mentioned above \citep{Namouni1999}.

The solar system contains several examples of co-orbital subsystems, for which there exists a very strong hierarchy between the masses of their components, as for the Jovian Trojans (or more generally of the giant planets), which correspond to a triplet  Sun/Jupiter/asteroid, or the systems of the Trojan satellites of Saturn, like the subsystem Saturn/Helene/Dione.
The only exception is the triplet Saturn/Janus/Epimetheus, where the masses of the two satellites differ only by a factor of three.

The detection of co-orbital exoplanets with current instrumentation is very challenging. Moreover, as co-orbital planets are expected to have neighbouring companions, their detection from observational data is a highly degenerate problem. 
In order to observe these systems, several techniques have been explored or developed, namely, radial velocity \citep{LeRoCo2015}, transits \citep{Janson2013Apj,HiAn2015}, combinations of transit and radial velocity \citep{FoGa2006,LeRoCoLi2017}, or transit timing variations of the planet \citep{FoHo2007,MaWi2009,VoNe2014}. 
Nonetheless, despite several studies dedicated to this quest \citep{MaWi2009,Janson2013Apj,LiBaFiLeSaCoRoFa2018,LiLePaFietal2018}, no co-orbital planets have been detected so far.

The theories of planetary formation do not prohibit the existence of co-orbital planets.  Indeed, \cite{LauCha2002}  introduced two possible processes that can form such systems: planet-planet gravitational scattering, and accretion in situ at the stable Lagrange points of a primary.
Depending on the physical characteristics of the gas disc, the first process can lead to systems with a high diversity of mass ratios \citep{CreNe2008}, but also to equal mass co-orbitals \citep{GiuBe2012}. 
 In the in-situ scenario, different models lead to various upper limits to the mass that can form at Lagrange's equilibrium points of a giant planet: a maximum mass of about 0.6 Earth mass for \cite{Beauge_etal_2007}, while \cite{LyJo2009} obtained 5 to 15 Earth mass planets at the same locations.  
 Once formed in the disc, the stability of the co-orbital system is not necessarily guaranteed. \cite{Beauge_etal_2007} found that inward migration tends to slightly increase the libration amplitude of the co-orbital system, and instability during the late migrating stages with low gas friction may lead to the destruction of the system. Another study from \cite{PiRya2014} shows that equal mass co-orbitals (from super-Earths to Saturns) are heavily disturbed during the gap-opening phase of their evolution.
Recently, \cite{LeCoAt2019} studied the dynamics of a pair of migrating co-orbital planets and showed that, depending on the mass-ratio, the eccentricities of the planets and the type of dissipative forces, the two planets may evolve towards the stable Lagrangian points, or may scatter out of the system. 
For systems with orbital periods $P_{\rm orb} \lesssim 10$~days, the planets undergo strong tidal interactions with the parent star \citep[e.g.,][]{Correia_etal_2020}, which arise from differential and inelastic deformation of the planet. 
In the two-body problem, the ultimate stage for tidal evolution is the synchronization of the rotation and orbital periods, alignment of the planet spin axis with the normal to the orbit, and the circularization of the orbit \citep[e.g.,][]{Hut_1980, Adams_Bloch_2015}.
In the full \fat{N}-body problem, 
\citet{Moeckel2017} proved that a relative equilibrium (solid rotation of the whole configuration) is never an energy minimiser of the space phase at a given total angular momentum. Thus, applied to the three-body problem with tidal dissipation, this result implies that the Lagrangian equilibria are made unstable by tides. 
However, we know neither the timescale of such instability nor if the phenomenon expands to the whole space phase, and we even less know what are the consequences to the dynamics of the co-orbital configuration.  
Indeed, although the spin of close-in planets quickly evolves into an equilibrium configuration, the orbital evolution is much slower \citep[e.g.,][]{Correia_Laskar_2010B}, and so the co-orbital configuration may survive the whole age of the system.
\citet{RodriguezGiupponeMichtchenko2013} performed some numerical simulations in the case of two identical co-orbital planets, but so far no analytical results have been provided to the consequences of tidal dissipation in the co-orbital resonance.

The present work attempts to study the influence of tidal dissipation in the dynamics of co-orbital planets.  
We develop an analytical model to account for tidal effects on the rotation, eccentricities and libration amplitude of these bodies.
We consider only the tidal effects raised by the star on the planets, which dominate the tidal evolution.
We thus obtain the non-conservative effects that shape the long-term dynamics. In section \ref{conservative}, we summarize the main known results on the co-orbital unrestricted three-body problem in the planar case and we introduce the elements useful to the following sections. In section \ref{sec:tide}, we establish the tidal formalism and we give analytic answers to the aforementioned questions in the vicinity of the Lagrangian points. We also present a tool to help with the detection of co-orbital exoplanets in section \ref{tool}. The results from section \ref{sec:tide} are expanded to the whole space phase in section \ref{S4}, where the validity of section \ref{sec:tide} is also tested numerically. 
In table \ref{notation} of Appendix~\ref{append_notation}, we list all the notations used throughout this paper.

\section{The co-orbital resonance}
\label{conservative}

\subsection{The averaged Hamiltonian}
\label{sec:averaged}
%
In order to construct the Hamiltonian associated with the coorbital resonance, we only consider in this section the point mass planar planetary three-body problem. The case of extended bodies is studied in Section \ref{sec:tide}. 
We start with two planets of masses  $m_1$ and $m_2$ small with respect to the mass $m_0$ of the star around which they orbit. 
For both planets we define the quantities $\beta_j = m_0m_j/(m_0+m_j)$, $\mu_j=\mathcal{G}(m_0+m_j),$ where $\mathcal{G}$ is the gravitational constant.  We also introduce the small parameter $\eps = (m_1 + m_2)/m_0$. 
In order to define a canonical coordinate system related to the elliptic elements $(a_j, e_j, \lambda_j, \varpi_j)$ (respectively the semi-major axis, the eccentricity, the mean longitude and longitude of the pericenter of the planet $j$), we start from Poincar\'e heliocentric coordinates  $(\lambda_j, \tilde{x}_j, \Lambda_j, x_j)$ where 
\be
\Lambda_j=\beta_j\sqrt{\mu_j a_j}, \quad x_j=\sqrt{\Lambda_j} \sqrt{ 1-\sqrt{1-e_j^2}}\exp(i\varpi_j), \quad \tilde{x}_j=-i\bar{x}_j.
\ee
In these coordinates, the Hamiltonian system derives from the sympletic form  
\be
\Omega = \sum_{j\in\{1,2\}} \left( d\lambda_j\wedge d\Lambda_j + d\tilde{x}_j \wedge dx_j \right).
\ee
Following \cite{RobutelPousse2013}, the planetary Hamiltonian can be written as
\be
H = H_K(\Lambda_1,\Lambda_2)+H_P(\Lambda_1,\Lambda_2,\lambda_1,\lambda_2,x_1,x_2,\tilde{x}_1,\tilde{x}_2),
\ee
where the Keplerian part $H_K$, of order $\eps$, takes the form 
\be
  H_K(\Lambda_1,\Lambda_2) = -\sum_{j =1}^2 \frac{\beta_j^3\mu_j^2}{2\Lambda_j^2}.
\ee
The perturbative part $H_P$, due to planet-planet interactions, takes into account both direct and indirect effects. It is a quantity of order $\eps^2$,  and can be expanded, at least for small to moderate eccentricities, in power series of $(x_j, \tilde{x}_j)$ with coefficients that are trigonometric polynomials in $\lambda_j$ depending on $\Lambda_j$.

We assume that the system is in, or close to, the co-orbital resonance. In this case, the angle $\lambda_1 - \lambda_2$ varies slowly with respect to the mean longitudes. We can therefore study the dynamics of the averaged problem over a fast angle. Moreover,  in the 1:1 mean-motion resonance, the values of the semi-major axes are always close to the same constant quantity denoted by $\ab$. As a consequence, the action variables $\Lambda_j$ will remain close to $\mathbf{\Lambda}^{\!\star}$ defined as
\be
\mathbf{\Lambda}^{\!\star} = (\Lambda_1^\star,\Lambda_2^\star)\, , \quad\text{with}\quad \Lambda_j^\star = m_j\sqrt{\mu_0\ab}\, ,
\quad\text{where}\quad
\mu_0 = \mathcal{G}m_0.
\ee
It follows that the mean-motion of the planet $j$ at  $\Lambda_j^\star$ satisfies
\be
\frac{d\lambda_j}{dt}=\drond{H_K}{\Lambda_j}(\mathbf{\Lambda}^{\!\star}) =\frac{\beta_j^3}{m_j^3}\frac{\mu_j^2}{\mu_0^{3/2}}\ab^{-3/2}= \eta \left(\accolade 1 + \gO(\eps) \accolade\right),
\ee
where the Kepler law reads 
\be
\eta=\sqrt{\mu_0\bar{a}^{-3}}.
\ee
Since we are only interested in a study of the problem in the vicinity of the resonance, we expand the Hamiltonian in a neighborhood of $(\Lambda_1,\Lambda_2) = (\Lambda_1^\star,\Lambda_2^\star)$ defined above. Then, in order to average the Hamiltonian, we perform the following variable transformation
\be\label{change_var_1}
(\Lambda_1,\Lambda_2,\lambda_1,\lambda_2) \longmapsto (Z,Z_2,\phi,\phi_2) = (\Lambda_1-\Lambda_1^\star, \, \Lambda_1+\Lambda_2 - \Lambda_1^\star -\Lambda_2^\star,\, \lambda_1-\lambda_2,\, \lambda_2).
\ee
As described by \cite{NiedermanPousseRobutel2020}, if we constrain the $\Lambda_j$ to belong to a neighborhood of $\mathbf{\Lambda}^{\!\star}$ of order $\eps^{1+\iota}$ with $1/2\geq \iota >1/3$,  the Keplerian part reads
\begin{equation}
\begin{aligned}
H_K(\Lambda_1,\Lambda_2) & =  \eta Z_2-\frac{3}{2}\eta\left(\frac{Z^2}{\Lambda_1^\star}+\frac{\left(Z_2-Z\right)^2}{\Lambda_2^\star}\right)   + \, \gO\left(\eps^{\iota+2}\right) \\
& = \hat{H}_K(Z,Z_ 2) + \gO\left(\eps^{\iota+2}\right)\, ,
\end{aligned}
\label{eq:H_K}
\end{equation}
where the constant terms have been neglected.

Since the perturbation is of order $\eps^2$, a zero-order expansion in $\Lambda_j - \Lambda_j^\star$ generates a remainder that has the same size as in  (\ref{eq:H_K}). Thus, we will limit ourselves to
\be
\begin{split}
&H_P(\Lambda_1,\Lambda_2,\lambda_1,\lambda_2,x_1,x_2,\tilde{x}_1,\tilde{x}_2)= H_P(\Lambda_1^\star,\Lambda_2^\star,\lambda_1,\lambda_2,x_1,x_2,\tilde{x}_1,\tilde{x}_2) +\,  \gO\left(\eps^{\iota+2}\right) \\
&\place\place\place\place\place\place\place\place\place\place=\hat{H}_P(\phi,\phi_2,x_1,x_2,\tilde{x}_1,\tilde{x}_2)+\gO\left(\eps^{\iota+2}\right).
\end{split}
\label{eq:H_P}
\ee 

In order to uncouple the variables associated with the different timescales, we perform the following linear transformation
\be\label{change_var_2}
(Z,Z_2,\phi,\phi_2) \longmapsto (I,I_2,\xi,\xi_2) = (Z - \delta Z_2, \, Z_2,\, \phi,\, \delta\phi + \phi_2),
\quad\text{with}\quad \delta=\frac{m_1}{m_1+m_2}, 
\ee
where \fat{x_j} and \fat{\tilde{x}_j} are unchanged. 
Thus, in the $(I,I_2)$ variables, the Keplerian part  $ \hat{H}$ takes the following form
\be
\hat{H}_K(Z,Z_2) =  \check{H}_K(I,I_2) = \eta I_2 - \frac32\eta\frac{I_2^2}{\Lambda_1^\star + \Lambda_2^\star}
 -  \frac32\eta\frac{\Lambda_1^\star + \Lambda_2^\star}{\Lambda_1^\star \Lambda_2^\star} I^2  .
\ee
This decoupling stresses the different dynamical timescales involved in the problem. Indeed, as
\be
\dot\xi_2 = \drond{\check{H}_K}{I_2} = \eta + \gO(\eps^\iota)
\quad \text{and}\quad 
\dot\xi = \drond{\check{H}_K}{I} = -3\eta\frac{\Lambda_1^\star + \Lambda_2^\star}{\Lambda_1^\star \Lambda_2^\star} I = \gO(\eps^\iota),
\ee 
a fast motion related to $\eta$ is associated with the orbital angle $\xi_2$, a semi-fast motion is associated to the resonant angle $\xi$, while a slow (or secular) evolution corresponds to the evolution of the variables $x_j$ (see below).

It is therefore legitimate to average the Hamiltonian over the fast angle $\xi_2$. For details of the averaging process, we refer to \cite{RobutelPousse2013} and  \cite{NiedermanPousseRobutel2020}.  Simply remember that replacing the perturbation 
\be
\check{H}_P(\phi,\phi_2,x_1,x_2,\tilde{x}_1,\tilde{x}_2) = \hat{H}_P(\xi,\xi_2,x_1,x_2,\tilde{x}_1,\tilde{x}_2)
\ee
by its average over the fast angle $\xi_2$ amounts to neglecting a remainder of order $\eps^{\iota+2}$. This last statement remains valid as long as the distance between the two planets does not go towards zero with $\eps$ \citep[see][for more details]{RoNiPo2016}.  Finally, we are left with the averaged Hamiltonian 
\be
\check{H}(I,I_2,\xi,x_1,x_2,\tilde{x}_1,\tilde{x}_2) = \check{H}_K(I,I_2) + \frac{1}{2\pi}\int_0^{2\pi}\check{H}_P(\xi,\xi_2,x_1,x_2,\tilde{x}_1,\tilde{x}_2) d\xi_2  .
\ee
Since $\xi_2$ no longer appears in the Hamiltonian, its conjugated action, $I_2 =  \Lambda_1+\Lambda_2 - \Lambda_1^\star -\Lambda_2^\star$, is an integral of the averaged Hamiltonian $\check{H}$.

Our last transformation consists in introducing dimensionless variables that are no longer proportional to the planetary masses (or to their square root).
We thus define the new coordinates $(J,J_2,X_1,X_2,\Xb_1,\Xb_2)$ as
\begin{equation}
J = \frac{I}{m\ab^2\eta},\quad 
J_2 = \frac{I_2}{m\ab^2\eta}, \quad
X_j = \sqrt{\frac{2}{m_j\ab^2\eta}}\,x_j, \quad
\Xb_j = i\sqrt{\frac{2}{m_j\ab^2\eta}}\,\tilde{x}_j ,
\label{change_var_3}
\end{equation}
where $m=\sqrt{m_1m_2}$.  
$J$ and $J_2$ are at most of order $\eps^{1/2}$ for tadpole orbits and $\eps^{1/3}$ for horseshoe one \cite[see][]{RobutelPousse2013}, while $X_j$ is close to the eccentricity vector of the planet $J$, that is
 \begin{equation}
X_j = e_j \exp(i\varpi_j)\left(\accolade 1 + \gO\left(J\right) + \gO\left(J_2\right) + \gO(\eps)  + \gO\left(e_j^2\right) \accolade\right) .
\end{equation}
The semi-major axes \fat{a_j} are linked to the variables \fat{J} and \fat{J_2} by the relations
\begin{equation}
a_j=\bar{a}\R_j^2,
\end{equation}
with
\begin{equation}
\R_j=1+f_j,\place f_1=\frac{m}{m_1+m_2}J_2+\frac{m}{m_1}J,\place f_2=\frac{m}{m_1+m_2}J_2-\frac{m}{m_2}J .
\end{equation}
In order to make the Hamiltonian equations as close as possible to their standard form, we rescale both time and energy as
\be
\Ham = \frac{\check{H}}{m\ab^2\eta^2} \quad \text{and}\quad  \tau = t\eta.
\ee
The equations of the motion become
\be
\begin{aligned}
&\dot{J} = -\drond{\Ham}{\xi}, \quad \dot{J_2} = -\drond{\Ham}{\xi_2},    
&\dot{\xi} = \drond{\Ham}{J},  \quad \dot{\xi_2} = \drond{\Ham}{J_2}, \\
&\dot{X}_j=-2i\frac{m}{m_j}\drond{\Ham}{\bar{X}_j}, 
&\dot{\bar{X}}_j=2i\frac{m}{m_j}\drond{\Ham}{X_j}.
\end{aligned}
\label{hamiltonX}
\ee
where the dot denotes the derivation with respect to $\tau$.
The Hamiltonian $\Ham$ now reads
\begin{equation}\label{eq:H_moy}
\Ham = \Ham_K + \Ham_P, \quad \text{with}\quad \Ham_K(J,J_2) = -\frac{3}{2}\frac{m_1+m_2}{m}J^2-\frac{3}{2}\frac{m}{m_1+m_2}J_2^2+J_2.
\end{equation}
As mentioned above, the perturbation $\Ham_P$ can be expanded in powers of the $(X_j, \Xb_j)$ as
\be
\Ham_P = \sum_{n\geq0} \Ham_{2n} \quad\text{where} \quad
\Ham_{2n} = \sum_{\vert\vect{p}\vert=2n} \Psi_{\vect{p}}\left(\Delta^{-1},e^{i\xi},e^{-i\xi}\right) X_1^{p_1}X_2^{p_2}\Xb_1^{\bar{p}_1}\Xb_2^{\bar{p}_2} \, .
\ee
In the previous expression, $\vect{p}$ is the multi-index $(p_1,p_2,\bar{p}_1,\bar{p}_2) \in \mathbb{N}^4$,  $\Psi_\vect{p}$ is a polynomial in \\ $\left(\Delta^{-1},e^{i\xi},e^{-i\xi}\right)$ and $\Delta = \sqrt{2 - 2\cos\xi}$.
The invariance by rotation of the system, which is equivalent to the conservation of its total angular momentum, yields an additional constraint on the multi-indexes $\vect{p}$ known as the D'Alembert rule
\be
p_1+p_2  = \bar{p}_1 + \bar{p}_2 \, .
\label{dalembert}
\ee
The expression of $\Ham_P$ expanded up to degree $4$ reads
\begin{equation}
\begin{split}
&\Ham_0=\frac{m}{m_0}\left(\cos\xi - \left(2-2\cos\xi\right)^{-1/2} \right), \\
&\Ham_2=\frac{1}{2}\frac{m}{m_0}\left\lbrace\accolade A_h\left( X_1\bar{X}_1+X_2\bar{X}_2 \right)+B_hX_1\bar{X}_2+\bar{B}_h\bar{X}_1X_2 \accolade\right\rbrace, \\
&\Ham_4=\frac{1}{4}\frac{m}{m_0}\left\lbrace\accolade D_h\left(X_1^2\bar{X}_1^2+X_2^2\bar{X}_2^2\right)+E_hX_1^2\bar{X}_2^2+\bar{E}_hX_2^2\bar{X}_1^2\right .\\ 
& \left . +F_h\left( X_1X_2\bar{X}_1^2+\bar{X}_1\bar{X}_2X_2^2\right) +\bar{F}_h\left(\bar{X}_1\bar{X}_2X_1^2+X_1X_2\bar{X}_2^2\right) +G_hX_1X_2\bar{X}_1\bar{X}_2\accolade\right\rbrace  ,
\end{split}
\label{eq:H_moy_detail}
\end{equation}
with
\begin{equation}
A_h=\frac{5\cos 2\xi-13+8 \cos \xi}{4\Delta^5}-\cos \xi \, , \quad
B_h=e^{-2i\xi}-\frac{ e^{-3i\xi}+16e^{-2i\xi}-26e^{-i\xi}+9e^{i\xi} }{8\Delta^5}.
\label{Ah_conservative}
\end{equation}
The coefficients \fat{D_h,\,E_h,\,F_h} and \fat{G_h} of the fourth degree are given in Appendix \ref{append_H4}.

\subsection{The semi-fast dynamics}
\label{conservative_circular}
\begin{figure}[h]
	\centering
	\includegraphics[width=0.9\linewidth]{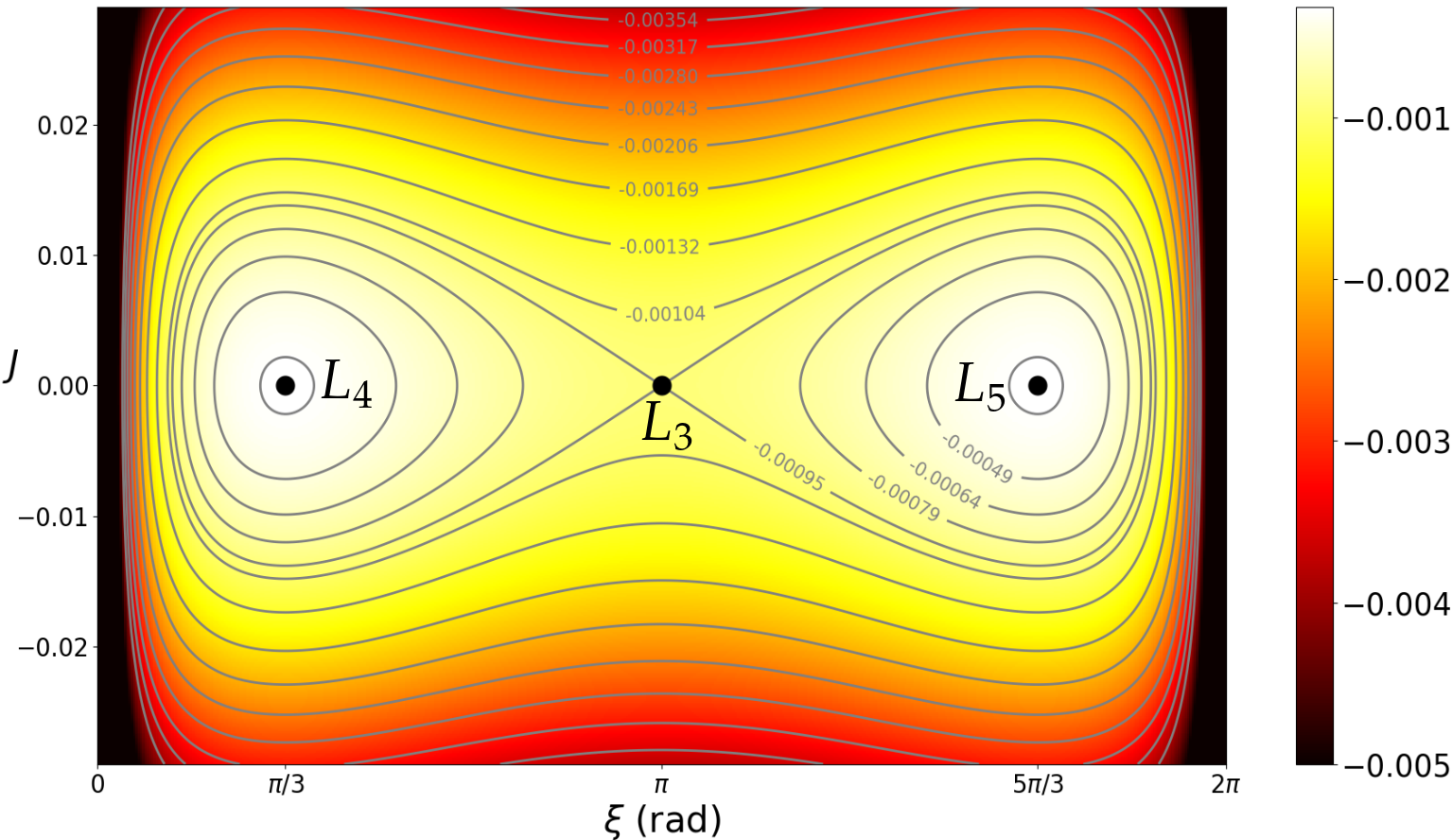}
	\caption{Phase portrait of the Hamiltonian \fat{\Ham_K+\Ham_0} with \fat{m_1=0.001\,m_0} and \fat{m_2=0.0004\,m_0}. The two elliptic fixed points \fat{L_4} and \fat{L_5} are energy maximizers while the hyperbolic fixed point \fat{L_3} is on a saddle point. Around \fat{L_{4,5}} are the tadpoles orbits and outside of the separatrix emanating from \fat{L_3} are the horseshoe orbits. The Hamiltonian tends to minus infinity as the angular separation between the planets $\xi$ moves  towards zero. 
}\label{space_phase}
\end{figure}

Since the Hamiltonian is even in the variables $X_j$ and $\Xb_j$, the manifold $X_1 = X_2 =0$ is invariant by the flow of the averaged Hamiltonian $\Ham$. In other words, the average problem has solutions for which both orbits are circular, but not necessarily Keplerian. Indeed, along such motions, the eccentricities remain equal to zero while the semi-major axes evolve. The dynamics of these particular trajectories  derives from the one degree of freedom Hamiltonian $\Ham_K+\Ham_0$  whose phase portrait is plotted in Figure \ref{space_phase}.
This Hamiltonian possesses three fixed points located in \fat{\left(J=0,\xi=\pi/3\right)}, \fat{\left(J=0,\xi=5\pi/3\right)} and \fat{\left(J=0,\xi=\pi\right)} corresponding to the famous Lagrangian equilibria. The two stable points that correspond to configurations where the three bodies occupy the vertices of an equilateral triangle are labeled by $L_4$ and $L_5$, while aligned Euler configuration is denoted by $L_3$.
The separatrices emanating from the unstable equilibrium point $L_3$ split the phase space in three distinct regions. The first two areas  contain the tadpole orbits inside the two lobs including respectively $L_4$ and $L_5$.  These trajectories, that surround the stable equilibria, are periodic with a frequency of order $\sqrt\eps$ (and equal to $\sqrt{27\eps/4}$ at $L_4$ or $L_5$). The third domain corresponds  to the outer region where the orbits called horseshoe surround the three fixed points. 
The $L_1$ and $L_2$  points are absent from this phase portrait because of the zero order truncation performed to the perturbation. They are replaced, in this model, by the singular line $\xi=0$ where the Hamiltonian $\Ham_K + \Ham_0$ tends to minus infinity, regardless of the values of the action $J$.  As a result, the model that we consider here is not valid for very large amplitude horseshoe orbits \citep[see][]{RobutelPousse2013}.  It is also worth mentioning that the libration frequency (semi-fast frequency) tends towards infinity when approaching the singularity and the averaging process no longer makes sense in the neighbourhood of the singularity.

\subsection{The secular dynamics}
\label{conservative_excentrique}

As we will show in Section \ref{sec:tide}, tidal dissipations tend to circularize the orbits. Thus, in this section, we will limit the study to the orbits with small eccentricities. For that purpose, it is sufficient to truncate the Hamiltonian equations to the first order in the $X_j$ along the circular orbits described above (Sect. \ref{conservative_circular}), which is equivalent to considering the trajectories associated to the Hamiltonian $\Ham_K + \Ham_0 + \Ham_2$.

According to (\ref{eq:H_moy_detail}), the variational equations are given by 
\be
\begin{pmatrix}
\dot{X}_1\\ 
\dot{X}_2
\end{pmatrix}
 =
 -\frac{i}{m_0}
 \begin{pmatrix}
m_2A_h(\xi(t))   & m_2\bar{B}_h(\xi(t)) \vspace{1mm}\\
m_1B_h(\xi(t))   & m_1A_h(\xi(t))\
\end{pmatrix}
 \begin{pmatrix}
 X_1\\
 X_2
 \end{pmatrix} ,
\label{eq:limX}
\ee
where $(J(t), \xi(t))$ is a solution of the semi-fast Hamiltonian system derived from $\Ham_K +\Ham_0$. Despite being linear, the differential equation (\ref{eq:limX}) is, a priori, not integrable, except in some special cases.
Indeed, if one chooses a stationary solution corresponding to one of the fixed points of the semi-fast system, the linear equation (\ref{eq:limX}) is no longer time-dependent and therefore becomes trivially integrable. 
This case is fully studied in  \cite{RobutelPousse2013}. Here, we just remind the main results that play an important role in the next sections.
We are not interested in the unstable configuration $L_3$ and we restrict the study to the stable equilibria $L_4$ and $L_5$.
For symmetry reasons, the dynamical features of the two equilateral configurations are the same,  therefore only the $L_4$ configuration ($\xi = \pi/3$) is studied.   
In this case,  the linear system (\ref{eq:limX}) has two purely imaginary eigenvalues
\be
ig_1 = i\frac{27}{8}\eps, \quad ig_2 = 0,
\label{eq:vp_secw}
\ee
and the associated eigenvectors are given by
\be
V_1 = 
\begin{pmatrix}
  m_2e^{i\frac{\pi}{3}} \\ 
  -m_1
\end{pmatrix}
\quad\text{and} \quad 
V_2 = 
\begin{pmatrix}
  e^{i\frac{\pi}{3}} \\ 
  1
\end{pmatrix} \, .\ee
The physical meaning of these eigenvectors can be interpreted as follow. Along the neutral direction, collinear to $V_2$, the elliptic elements of both orbits satisfy the relations
\be\label{pisur3}
\frac{X_1}{X_2} = \exp\left(i\frac{\pi}{3}\right), 
\ee
a configuration that corresponds to the Lagrange elliptic equilibrium, which is a stable fixed point of the average problem.  Along the other direction, collinear to $V_1$, the two planets are linked (at least for small eccentricities) by the relations
\be\label{4pisur3}
 \frac{X_1}{X_2} = \frac{m_2}{m_1}\exp\left(i\frac{4\pi}{3}\right).
 \ee
This corresponds to an infinitesimal version of the anti-Lagrange orbits found numerically by \cite{Guippone2010}. These particular orbits are remarkable in the sense that, in the average problem, $a_1, a_2, e_1, e_2$ and $\lambda_1 - \lambda_2$ are constant, while the two orbits precess with the same frequency equal to $g_1$ in such a way that  the angle $\varpi_1 - \varpi_2$ is constant and equal to \fat{4\pi/3}.

If we now consider the neighbourhood of a purely periodic solution, as those plotted in Fig. \ref{space_phase}, the solutions of (\ref{eq:limX}) can be approximated by averaging over the semi-fast variations.  Indeed, the frequency of the semi-fast motion is of order $\sqrt\eps$, while the secular frequency is of order $\eps$ (eq. (\ref{eq:vp_secw})). Unfortunately, the averaged expression of the differential system (\ref{eq:limX}) cannot be given explicitly\footnote{\cite{NiedermanPousseRobutel2020} give an expression of the coefficients of this matrix in terms of integrals depending on some parameters.}. In the horseshoe domain, some information can nevertheless be obtained from geometrical considerations. Since \fat{\mathcal{I}_{\text{m}}\left(B_h(2\pi - \xi)\right) = -\mathcal{I}_{\text{m}}\left(B_h(\xi)\right)} and because horseshoe orbits are symmetric with respect to $(J,\xi)= (0,\pi)$, the average of \fat{\mathcal{I}_{\text{m}}\left(B_h\right)} over the semi-fast variations vanishes. As a result,  the coefficients of the averaged variational equation are purely imaginary numbers. Taking into account the particular form of the matrix, it results that the two eigenvectors correspond to  pairs of orbits whose pericenters are aligned for the Lagrange configuration (\fat{\varpi_1-\varpi_2=0}) and anti-aligned for the anti-Lagrange configuration (\fat{\varpi_1-\varpi_2=\pi}) (Appendix \ref{append_hs} provides more details). Nothing similar is valid for tadpole orbits where algebraic calculations are required.

\section{Tidal effects}\label{S3}
\label{sec:tide}
%
%

\subsection{Tidal model}

Tides arise from differential and inelastic deformations of a body~\fat{j} (e.g., the planet) due to the gravitational effect of a perturbing body~\fat{i} (e.g., the star or the companion planet).
Tidal contributions to the orbital and spin evolution are based on a very general formulation of the tidal potential, initiated by \citet{Darwin_1880}. 
Since celestial bodies are not perfectly rigid, there is a distortion that gives rise to a tidal bulge. 
This redistribution of mass modifies the gravitational potential generated by the body~\fat{j} in any point of the space, \fat{\vect{r}}.
The additional amount of potential is known by the tidal potential \citep[e.g.][]{Kaula_1964}
\begin{equation}\label{perturbed}
V(\vect{r})=-\kappa_{2,j}\frac{\mathcal{G}m_i}{R_j}\left(\frac{R_j}{r}\right)^3\left(\frac{R_j}{r_i^{\bigstar}}\right)^3 P_2\left(\cos S\right) ,
\end{equation}
where the indice \fat{i} refers to the body responsible for the tidal bulge (perturbing body), while the indice \fat{j} stands for the body where the bulge is raised. 
\fat{\vect{r}_i} 
is the position of the perturbing body 
with respect to the barycenter of body \fat{j}, \fat{R_j} is the radius of the body~\fat{j}, $\kappa_{2,j}$ is the second Love number, \fat{P_2} is the second Legendre polynomial, and \fat{S} is the angle between \fat{\vect{r}_i^{\bigstar}} and \fat{\vect{r}}.
For an interacting body with mass $m_k$ at the position \fat{\vect{r}_k} the amount of tidal potential energy is then given by 
\begin{equation}\label{perturbed_energy}
U_{ijk} (\vect{r}_k) = m_k V(\vect{r}_k)=-\kappa_{2,j}\frac{\mathcal{G}m_k m_i}{R_j}\left(\frac{R_j}{r_k}\right)^3\left(\frac{R_j}{r_i^{\bigstar}}\right)^3 P_2\left(\cos S\right).
\end{equation}

A system of \fat{N} bodies undergoing tidal forces is described by \fat{N\left(N-1\right)^2} tidal potentials. For \fat{N=3}, this gives \fat{12} contributions, but tides raised on a planet by another planet, as well as tides raised by the central star on a planet and interacted with by the other planet can be neglected due to their very small contribution. 
Only \fat{6} contributions remain: \fat{4} of these are of order $\kappa_{2,0}(\varepsilon m_0)^2 R_0^5$, 
and correspond to tides raised on the star and interacted with by the planets, while the last \fat{2} are of order $\kappa_{2,j} m_0^2 R_j^5$, 
and correspond to tides raised on a planet by the star and interacted with by the star.
For planets in the solar system we usually have $\kappa_{2,j} \approx 0.5$ \citep{Yoder_1995cnt}, while for Sun-like stars we have $\kappa_{2,0} \approx 0.02$ \citep{Claret_Cunha_1997}.
Assuming a constant density for all bodies we have $R_j \propto m_j^{1/3}$ and thus
\begin{equation}\label{neglect_star}
\frac{\kappa_{2,0}(\varepsilon m_0)^2 R_0^5}{\kappa_{2,j} m_0^2 R_j^5} = \frac{\kappa_{2,0}} {\kappa_{2,j}} \varepsilon^{1/3} \ll 1 ,
\end{equation}
that is, tides raised on the star can also be neglected.
Therefore, in this work we consider only the 2 terms corresponding to tides raised on each planet by the star and interacted with by the star.

The dissipation of the mechanical energy inside the planet introduces a delay $\Delta t_j$, and hence a phase shift between the initial perturbation and the maximal tidal deformation.
As a consequence, the star exerts a torque on the tidal bulge which modifies the spin and the orbit of the planet.
In all the following, we use the notation
\begin{equation}\label{zt_deltat}
z_j^{\bigstar}\left(t\right) = z_j\left(t-\Delta t_j\right) , 
\end{equation}
\fat{z_j} being any quantity.

Tidal dissipation is usually modeled through the quality factor $Q_j$, which measures the amount of energy dissipated in a tidal cycle \citep[e.g.,][]{Munk_MacDonald_1960}.
For a given tidal frequency $\sigma$, the tidal dissipation can be related to this delay through \citep[e.g.,][]{Efroimsky_2012}
\be
Q_j^{-1} (\sigma) = \sin ( \sigma \Delta t_j (\sigma) ) \approx \sigma \Delta t_j (\sigma).
\label{171013b}
\ee
The exact dependence of $\Delta t_j (\sigma)$ on the tidal frequency is unknown.
In order to take into account tidal dissipation, we need to adopt a tidal model.
A large variety of models exists, but the most commonly used are the constant-$Q$ \citep[e.g.,][]{Munk_MacDonald_1960}, the linear model \citep[e.g.,][]{Mignard1979}, 
the Maxwell model \citep[e.g.,][]{Correia_etal_2014},
and the Andrade model \citep[e.g.,][]{Efroimsky_2012}.
Some models appear to be better suited to certain situations, but no model is globally accepted.
Nevertheless, regardless of the tidal model adopted, the qualitative conclusions are more or less unaffected, and the system always loses mechanical energy.

In this work we adopt a viscous linear model for tides \citep{Singer1968, Mignard1979}.
In this model it is assumed that the time delay $\Delta t_j $ is constant and independent of the frequency.
This tidal model is widely used and provides very simple expressions for the tidal interactions.

\subsection{The dynamical impact of tidal dissipations}\label{tidal_impact}

\subsubsection{Derivation of the equations of motion}

Although tidal effects do not preserve the energy, it is possible to use the Hamiltonian formalism by considering the starred quantities as parameters \citep{Mignard1979}.

The tidal Hamiltonian reads
\begin{equation}
H_t=H_t^1+H_t^2+T_1+T_2 ,
\label{eq:H_tide}
\end{equation}
where, in the heliocentric reference frame
\begin{equation}\label{Ht_cartesien}
H_t^j=-\kappa_{2,j}\mathcal{G}m_0^2\frac{R_j^5}{r_j^3r_j^{\bigstar3}}P_2\left(\cos S\right)
\quad\text{and} \quad T_j=\frac{\Theta_j^2}{2C_j},
\end{equation}
with
\begin{equation}
S=\lambda_j-\lambda_j^{\bigstar}-\left(\theta_j-\theta_j^{\bigstar}\right),
\end{equation}
where \fat{\theta_j} is the angle of rotation of body~\fat{j},
$\Theta_j = C_j\omega_j$ is the conjugated momentum of $\theta_j$, $C_j=\alpha_jm_jR_j^2$ is the moment of inertia of body $j$, $\omega_j=d\theta_j/dt$ is its rotation rate, and \fat{\alpha_j} is a dimensionless structure constant depending on the state equation of body \fat{j} (\fat{\alpha_j=2/5} for an homogeneous ball).

We use the canonical Poincar\'e rectangular variables to express the tidal Hamiltonian \fat{H_t} (see \cite{LaskarRobutel1995}) and perform the transformations given by equations (\ref{change_var_1}), (\ref{change_var_2}) and (\ref{change_var_3}), as well as the transformation
\begin{equation}
\Gamma_j=\frac{\Theta_j}{m\bar{a}^2\eta},
\end{equation}
in order to normalize the action variable associated with the angle \fat{\theta_j}.

Contrary to what was done for $\Ham_P$ in section \ref{sec:averaged}, it is not possible  to truncate the Hamiltonian (\ref{Ht_cartesien}) at the order \fat{0} in the variables \fat{J} and \fat{J_2} without losing relevant informations on the tidal dissipation.   Indeed, the main tidal contribution results from the fact that the semi-major axes are not constant but animated with semi-fast motions (see section \ref{conservative_circular}). As a consequence, we keep exact expressions of $\Ham_t^j$ in terms of $J$ and $J_2$, and obtain
%
%
%
\begin{equation}\label{Ht_rectangular}
\begin{split}
&\Ham_t^j=-q_j\frac{m_0}{m}\R_j^{-6}\R_j^{\bigstar-6}\left\lbrace\accolade A_t^j+\Xi_2^j+\Xi_4^j+\mathcal{O}\left(\left|X_j\right|^6\right) \accolade\right\rbrace,\\
&\mathcal{T}_j=\frac{1}{2\alpha_j}\frac{m}{m_j}\frac{\Gamma_j^2}{\qoppa_j^2}=\frac{T_j}{m\bar{a}^2\eta^2},
\end{split}
\end{equation}
with
\begin{equation}
\begin{split}
&\Xi_2^j=B_t^j\left(\R_j^{-1}X_j\bar{X}_j+\R_j^{\bigstar-1}X_j^{\bigstar}\bar{X}_j^{\bigstar} \right)+\left(\R_j\R_j^{\bigstar} \right)^{-1/2} \left( C_t^jX_j\bar{X}_j^{\bigstar}+\bar{C}_t^jX_j^{\bigstar}\bar{X}_j\right) , \\
&\Xi_4^j=D_t^j\left(\R_j^{-2}X_j^2\bar{X}_j^2+{{\R_j^{\bigstar-2}}}{X_j^{\bigstar2}}{\bar{X}_j^{\bigstar2}}\right)+\left(\R_j\R_j^{\bigstar} \right)^{-1}\left( E_t^jX_j^2\bar{X}_j^{\bigstar2}+\bar{E}_t^jX_j^{\bigstar2}\bar{X}_j^2\right) \\
&+\left(\R_j\R_j^{\bigstar} \right)^{-1}G_t^jX_jX_j^{\bigstar}\bar{X}_j\bar{X}_j^{\bigstar}+\R_j^{-3/2}\R_j^{\bigstar-1/2}\left( F_t^jX_jX_j^{\bigstar}\bar{X}_j^2+\bar{F}_t^jX_j^2\bar{X}_j\bar{X}_j^{\bigstar} \right)\\
&+\R_j^{-1/2}\R_j^{\bigstar-3/2}\left( F_t^jX_j^{\bigstar2}\bar{X}_j\bar{X}_j^{\bigstar}+\bar{F}_t^jX_jX_j^{\bigstar}\bar{X}_j^{\bigstar2} \right),
\end{split}
\end{equation}
where we defined
\begin{equation}
q_j=\kappa_{2,j}\qoppa_j^5,\place \qoppa_j=\frac{R_j}{\bar{a}},
\end{equation}
and
\begin{equation}
A_t^j=\frac{1}{4}+\frac{3}{4}\cos 2\left( \lambda_j-\lambda_j^{\bigstar}-\theta_j+\theta_j^{\bigstar} \right).
\end{equation}
The coefficients \fat{B_t^j} through \fat{G_t^j} of the second and fourth order in eccentricity are given in Appendix \ref{append_H4}.
The equations of motions are derived from the Hamiltonian leaving the starred variables constant since they are only parameters, that is:
\begin{equation}\label{scalar_to_vectorial}
\begin{split}
&\dot{X}_j=-2i\frac{m}{m_j}\drond{\Ham}{\bar{X}_j}\Bigr|_{\bigstar=\text{Cte}},\place \dot{\bar{X}}_j=2i\frac{m}{m_j}\drond{\Ham}{X_j}\Bigr|_{\bigstar=\text{Cte}},\\
&\dot{\vect P}=-\drond{\Ham}{\vect Q}\Bigr|_{\bigstar=\text{Cte}},\place\place\place\;\; \dot{\vect Q}=\drond{\Ham}{\vect P}\Bigr|_{\bigstar=\text{Cte}},
\end{split}
\end{equation}
where the upper dot denotes the derivation with respect to \fat{\tau} and
\begin{equation}
\vect P=\,^t\left(J,J_2,\Gamma_1,\Gamma_2\right),\place \vect Q=\,^t\left(\xi,\xi_2,\theta_1,\theta_2\right).
\end{equation}
The full Hamiltonian reads
\begin{equation}\label{Ht_planet}
\Ham=\Ham_K+\Ham_0+\Ham_2+\Ham_4+\sum_{j\in\left\lbrace1,2\right\rbrace}\Ham_t^j+\sum_{j\in\left\lbrace1,2\right\rbrace}\mathcal{T}_j.
\end{equation}
In the linear tidal model, once the equations of motion are obtained, the starred quantities $z_j^{\bigstar}$ are expressed as the first order Taylor expansion in \fat{\eta\Delta t_j} (Eq.\,(\ref{zt_deltat})) 
\begin{equation}\label{linear_model}
z_j^{\bigstar} = z_j - \eta\Delta t_j\dot{z}_j \ ,
\end{equation}
where \fat{\dot{z}_j} is expressed using equations (\ref{scalar_to_vectorial}). To prevent an implicit relation for \fat{z_j^{\bigstar}}, only the Keplerian and kinetic part of the Hamiltonian is considered for \fat{\dot{z}_j} in equation (\ref{linear_model}). That is, we take
\begin{equation}
\Ham=\Ham_K+\mathcal{T}_1+\mathcal{T}_2,
\end{equation}
in equation (\ref{scalar_to_vectorial}) only for the dot appearing in equation (\ref{linear_model}).
We redefine the quality factor of planet \fat{j} as (Eq.\,(\ref{171013b}))
\begin{equation}
Q_j=\frac{1}{\eta\Delta t_j}.
\end{equation}
If we note \fat{\vartheta_j=1-\frac{\omega_j}{\eta}}, the complete set of equations of motion reads
\begin{equation}\label{equation_case1}
\begin{split}
&\dot{\vartheta}_j=-3\alpha_j^{-1}\frac{m_0}{m_j}\qoppa_j^{-2}\frac{q_j}{Q_j}\R_j^{-12}\left\lbrace\accolade \vartheta_j+3\left(1-\R_j\right)+h_2^j\R_j^{-1}X_j\bar{X}_j+h_4^j\R_j^{-2}X_j^2\bar{X}_j^2  \accolade\right\rbrace,\\
&\dot{J}=-\drond{\left(\Ham_0+\Ham_2+\Ham_4\right)}{\xi}+\left(1-\delta\right)\dot{J}_2^1-\delta\dot{J}_2^2,\\ 
&\dot{J}_2=\dot{J}_2^1+\dot{J}_2^2,\\
&\dot{\xi}=\drond{\Ham_K}{J}+6q_1\frac{m_0}{m_1}\R_1^{-13}V_2\left(\R_1^{-1}X_1\bar{X}_1\right)-6q_2\frac{m_0}{m_2}\R_2^{-13}V_2\left(\R_2^{-1}X_2\bar{X}_2\right),\\
&\dot{X}_j=-2i\frac{m}{m_j}\drond{\left(\Ham_2+\Ham_4\right) }{\bar{X}_j}-3\frac{q_j}{Q_j}\frac{m_0}{m_j}\R_j^{-13}X_j\left\lbrace\accolade p_2^j-\frac{5}{2}iQ_j+\frac{X_j\bar{X}_j}{\R_j}\left(p_4^j-\frac{65}{4}iQ_j\right)\accolade\right\rbrace,
\end{split}
\end{equation}
where
\begin{equation}
\begin{split}
&\dot{J}_2^j=-3\frac{q_j}{Q_j}\frac{m_0}{m}\R_j^{-12}\left\lbrace\accolade \vartheta_j+3\left(1-\R_j\right)+k_2^j\R_j^{-1}X_j\bar{X}_j+k_4^j\R_j^{-2}X_j^2\bar{X}_j^2  \accolade\right\rbrace,\\
&h_2^j=\frac{93}{2}+\frac{15}{2}\vartheta_j-\frac{81}{2}\R_j,\;\;\;\;\;\;\;\;\;\;\;\;\;\;\;\;\;\;\;h_4^j=\frac{1989}{8}+\frac{195}{8}\vartheta_j-\frac{819}{4}\R_j,\\
&k_2^j=\frac{157}{2}+\frac{27}{2}\vartheta_j-69\R_j,\;\;\;\;\;\;\;\;\;\;\;\;\;\;\;\;\;\;k_4^j=\frac{2515}{4}+\frac{273}{4}\vartheta_j-\frac{2091}{4}\R_j,\\
&p_2^j=32+6\,\vartheta_j-\frac{57}{2}\R_j,\;\;\;\;\;\;\;\;\;\;\;\;\;\;\;\;\;\;\;\;\;\;p_4^j=\frac{3041}{8}+\frac{351}{8}\vartheta_j-318\R_j,\\
&V_2\left(Z\right)=1+\frac{65}{8}Z+\frac{455}{16}Z^2.\\
\end{split}
\end{equation}
In appendix \ref{append_angular} we show that this set of equations preserves the total angular momentum of the system. 

\subsubsection{Fixed points and linearization of the system}\label{linearization}

Since the system (\ref{equation_case1}) is a perturbation of the Hamiltonian system derived from (\ref{eq:H_moy}), its fixed points are a perturbation of the Lagrangian equilibrium. We can thus find them using a perturbative approach. Since system (\ref{equation_case1}) was written at first order in \fat{\varepsilon^{-1}\left(q_1+q_2\right)}, we restrict to the first order to compute the position of the fixed points. We have, for \fat{j\in\left\lbrace1,2\right\rbrace},
%

%
\begin{equation}\label{point_fixe_case1}
\begin{split}
\quad \vartheta_j&=0,\\
\xi-\frac{\pi}{3}&=0,\\
-3\frac{m_1+m_2}{m}J+6q_1\frac{m_0}{m_1}-6q_2\frac{m_0}{m_2}&=0,\\
X_j&=0.
\end{split}
\end{equation}
Note that this choice for the fixed points does not make the right hand side of system (\ref{equation_case1}) to be exactly zero but only a quantity of second order in \fat{\varepsilon^{-1}\left(q_1+q_2\right)}. This choice guarantees though that the fixed points correspond to a solid rotation of the whole system. We now fix to \fat{0} the value of \fat{\dot{\vartheta}_j} in (\ref{equation_case1}) and if we approximate \fat{a_j=\bar{a}} (that is \fat{\R_j=1}), the equilibrium of the rotation rate of planet \fat{j} at non-zero eccentricity reads
\begin{equation}\label{super_synchrone}
\frac{\omega_j}{\eta}=1+6e_j^2+\frac{3}{8}e_j^4+\mathcal{O}\left(e_j^6\right),
\end{equation}
which is known as the pseudo-synchronization \citep{Correia_Laskar_2004}. 
Indeed, this is not a solid rotation, because the equilibrium is slightly super-synchronous at non-zero eccentricities.

Equations (\ref{point_fixe_case1}) only provide 6 independent equations for 7 variables\footnote{The remaining degree of freedom is due to the conservation of the total angular momentum of the system.}. In order to have a unique fixed point in the neighbourhood of which to linearize the system, we arbitrarily choose
\begin{equation}\label{point_fixe_case1_arbitraire}
f_1+f_2=0.
\end{equation}
Note that the linearized system depends on the choice of both equation (\ref{point_fixe_case1}) and equation (\ref{point_fixe_case1_arbitraire}), but only by a quantity of second order in \fat{\varepsilon^{-1}\left(q_1+q_2\right)}, which we neglect.

Let \fat{\mathcal{X}_0=\,^t\left(\vartheta_{1,0},\vartheta_{2,0},\xi_0,J_0,J_{2,0},X_{1,0},X_{2,0}\right)} be the unique solution of (\ref{point_fixe_case1}) \fat{+} (\ref{point_fixe_case1_arbitraire}). In order to study the dynamics of the system in the neighbourhood of the fixed point, we consider the linearized system
\begin{equation}\label{linear_system}
\dot{\mathcal{X}}=\left(\mathcal{Q}_0+\mathcal{Q}_1\right)\mathcal{X},
\end{equation}
where \fat{\mathcal{X}=\,^t\left(\vartheta_1,\vartheta_2,\xi,J,J_2,X_1,X_2\right)-\mathcal{X}_0}. The matrix \fat{\mathcal{Q}_0} derives from the conservative Hamiltonian \fat{\Ham_K+\Ham_0+\Ham_2} (the fourth order in eccentricity does not contribute to the linearized system) while \fat{\mathcal{Q}_1} corresponds to the tidal contributions. The matrix \fat{\mathcal{Q}_0+\mathcal{Q}_1} is block diagonal with a size \fat{5} block corresponding to the circular dynamics and a size \fat{2} block, corresponding to \fat{X_1} and \fat{X_2}. This shows that the circular and eccentric dynamics are uncoupled near \fat{L_{4,5}}. A detailed expression of theses matrices is given in appendix \ref{append_linear}. The set of eigenvalues of \fat{\mathcal{Q}_0} is
\begin{equation}\label{SpQ0}
\left\lbrace 0,0,0,i\nu,-i\nu,ig_1,ig_2\right\rbrace ,
\end{equation}
where 
\begin{equation}
\nu=\sqrt{\frac{27\eps}{4}},\place g_1=\frac{27\varepsilon}{8},\place g_2=0.
\end{equation}
Among the three\footnote{Four, with \fat{g_2}} \fat{0} eigenvalues, two correspond to the constant rotation rate of the planets and another to the conservation of the total angular momentum. The \fat{\pm i\nu} eigenvalues give the frequency of the libration of the resonant angle \fat{\xi=\lambda_1-\lambda_2} around \fat{L_{4,5}} (see section \ref{conservative_circular}) and the last two eigenvalues give the frequency of the precession of the pericenters in the eigen-modes anti-Lagrange for \fat{g_1} and Lagrange for \fat{g_2} (see section \ref{conservative_excentrique}). In particular, all seven eigenvalues are pure imaginary and without the contribution of tides (\fat{\mathcal{Q}_1}), the trajectories of the linearized system are quasi-periodic.

Since \fat{\mathcal{Q}_1} is only a small perturbation of \fat{\mathcal{Q}_0}, we expect that the spectrum of \fat{\mathcal{Q}_0+\mathcal{Q}_1} is close to (\ref{SpQ0}). We compute it to the first order in \fat{\varepsilon^{-1}\left(q_1+q_2\right)} using results from appendix \ref{append_diago}. We find for the eigenvalues of \fat{\mathcal{Q}_0+\mathcal{Q}_1}
\begin{equation}\label{spectrum}
\left\lbrace\lambda_1,\lambda_2,0,\sampi,\bar{\sampi},\lambda_{\text{AL}},\lambda_{\text{L}}\right\rbrace,
\end{equation}
with
\begin{equation}\label{eigenvalues_circular}
\begin{split}
&\lambda_j=\boxed{-3\alpha_j^{-1}\frac{q_j}{Q_j}\qoppa_j^{-2}\frac{m_0}{m_j}+9\varepsilon^{-1}\frac{q_j}{Q_j}}<0, \\
%
%
&\sampi =  \boxed{\frac{9}{2}\varepsilon^{-1}\!\left(\frac{m_1}{m_2}\frac{q_2}{Q_2}+\frac{m_2}{m_1}\frac{q_1}{Q_1}\right)} + i\nu\left[1+13\varepsilon^{-1}\left(\frac{m_1}{m_2}q_2+\frac{m_2}{m_1}q_1\right)\right],\\
&\lambda_{\text{AL}}=\boxed{-\frac{21}{2}\varepsilon^{-1}\left(\frac{m_1}{m_2}\frac{q_2}{Q_2}+\frac{m_2}{m_1}\frac{q_1}{Q_1}\right)}+ig_1\left[1+\frac{20}{9}\varepsilon^{-2}\left(\frac{m_1}{m_2}q_2+\frac{m_2}{m_1}q_1\right)\right] , \\
&\lambda_{\text{L}}=\boxed{-\frac{21}{2}\varepsilon^{-1}\left(\frac{q_1}{Q_1}+\frac{q_2}{Q_2}\right)}+\frac{15}{2}i\varepsilon^{-1}\left(q_1+q_2\right).
\end{split}
\end{equation}
We note that the eigenvalues are no longer pure imaginary and we boxed the real parts for a better visualization. The real parts are proportional to the quantities \fat{Q_j^{-1}}, while the perturbations of the imaginary parts are not. As a consequence, elastic tides do not yield dissipation, but only change slightly the fundamental frequencies of the system. Tides also slightly perturb the eigenvectors of the system and we show in appendix \ref{append_linear} how the eigen-modes Lagrange and anti-Lagrange are modified by tides.

One of the eigenvalues of \fat{\mathcal{Q}_0+\mathcal{Q}_1} is still zero, corresponding to the conservation of the total angular momentum. %
%
%
As a result of tidal dissipation, the eigenvalues $\lambda_{\text{AL}}$ and $\lambda_{\text{L}}$, perturbations of \fat{ig_1} and \fat{ig_2}, respectively, have non-zero negative real parts. Therefore, both eccentric eigen-modes Lagrange and anti-Lagrange are damped to zero. Similarly, the rotation rates of both planets are also damped. On the contrary, the real part of  $\sampi$ and $\bar{\sampi}$, perturbations of $i\nu$ and $-i\nu$, are strictly positive, which leads to an exponential increase of the libration amplitude when the system is around \fat{L_{4,5}}.

\subsubsection{Characteristic timescales}\label{tidal_times}
We define here the characteristic timescale of a given proper mode of the system (\ref{linear_system}) as the time needed for its amplitude to be multiplied (or divided, if the corresponding real part is negative) by a factor \fat{\exp\left(1\right)}. According to the eigenvalues (\ref{eigenvalues_circular}), these times are
%
\begin{equation}\label{timescales}
\begin{split}
&\tau_{\text{rot}}^j=\frac{1}{6\pi}\alpha_j\qoppa_j^2\frac{m_j}{m_0}\frac{Q_j}{q_j}\left[1+\frac{3m_j}{m_1+m_2}\qoppa_j^2\alpha_j\right]T,\\
&\tau_{\text{L}}=\frac{\varepsilon}{21\pi}\left(\frac{q_1}{Q_1}+\frac{q_2}{Q_2}\right)^{-1}T,\\
&\tau_{\text{AL}}=\frac{\varepsilon}{21\pi}\left(\frac{m_2}{m_1}\frac{q_1}{Q_1}+\frac{m_1}{m_2}\frac{q_2}{Q_2}\right)^{-1}T,\\
&\tau_{\text{lib}}=\frac{7}{3}\tau_{\text{AL}},
\end{split}
\end{equation}
where \fat{T=2\pi/\eta} is the orbital period. We note that the times \fat{\tau_{\text{rot}}^j} are much smaller than the three other characteristic times, due to the presence of the factor \fat{\qoppa_j^2\ll1}. That is, regardless of the parameters and initial conditions, the rotations of the planets are damped to their equilibrium (\ref{super_synchrone}) in a timescale such that the eccentricities and the libration angle do not undergo significant damping or excitation.

We know from the eigenvalues (\ref{eigenvalues_circular}) that the two eccentric eigen-modes Lagrange and anti-Lagrange are damped to zero, while the libration amplitude of the resonant angle \fat{\xi} exponentially increases. We now compare the timescales \fat{\tau_{\text{AL}}}, \fat{\tau_{\text{L}}} and \fat{\tau_{\text{lib}}} to determine which proper mode amplitudes evolve faster. Even though both eccentric modes are damped, the damping times may be different. By comparing them, we can find if the system favours the Lagrange or the anti-Lagrange configuration. Indeed if
\begin{equation}
\begin{split}
&\frac{\tau_{\text{AL}}}{\tau_{\text{L}}}<1\;\rightarrow\;\text{ then the system settles in Lagrange whereas if},\\
&\frac{\tau_{\text{AL}}}{\tau_{\text{L}}}>1\;\rightarrow\;\text{ then the system settles in anti-Lagrange}.
\end{split}
\end{equation}
Moreover, comparing the time \fat{\tau_{\text{lib}}} with the eccentric times \fat{\tau_{\text{AL}}} and \fat{\tau_{\text{L}}} allows us to know if the system is still eccentric or already circular when the libration amplitude has significantly increased. We have
\begin{equation}\label{tau_lib}
\tau_{\text{lib}}=\frac{1}{9\pi}\frac{\varepsilon}{q_1/Q_1+q_2/Q_2}\frac{\tau_{\text{AL}}}{\tau_{\text{L}}}T,
\end{equation}
and so, for a given sum of the planetary masses, \fat{\varepsilon}, and sum of the dissipation rates, \fat{q_1/Q_1+q_2/Q_2}, the system moves away from \fat{L_{4,5}} faster if it favours Lagrange, and slower if it favours anti-Lagrange.

These results are a priori valid only in a small neighboorhoud of \fat{L_{4,5}}, but in fact, the simulations from section \ref{S4} show that the behaviour of the system near \fat{L_{4,5}} is valid even at high libration amplitudes. This means that the increase in the libration amplitude is unbounded, leading to a systematic destruction of the system due to close encounters between the planets. The disruption time depends on \fat{\tau_{\text{lib}}}, and so Lagrange-like systems have a short life expectancy, while anti-Lagrange-like systems have a long life expectancy. Moreover, a Lagrange-like system is eccentric when old\footnote{Old means that its age is significant with respect to its expectancy of life.} (as long as it was eccentric when young), while an anti-Lagrange-like system is always circular when old. Indeed, for an anti-Lagrange-like system, the characteristic time \fat{\tau_{\text{e}}} of eccentricity damping is given by \fat{\tau_{\text{AL}}} and then, \fat{\tau_{\text{lib}}=7\tau_{\text{e}}/3} ensures that the eccentricity is damped when the libration amplitude has significantly increased. On the other hand, for a Lagrange-like system, \fat{\tau_{\text{e}}=\tau_{\text{L}}} and then, \fat{\tau_{\text{lib}}\ll\tau_{\text{e}}} ensures that the system is still eccentric when the libration amplitude has significantly increased.

Although it is clear that \fat{\tau_{\text{lib}}} depends on the semi-major axis \fat{\bar{a}}, equation (\ref{tau_lib}) does not explicitly show it, since \fat{\bar{a}} is hidden in the variables \fat{q_j}, \fat{T} and \fat{Q_j}. We have
\begin{equation}
\tau_{\text{lib}}\propto\frac{\tau_{\text{AL}}}{\tau_{\text{L}}}\bar{a}^{\beta},
\end{equation}
where \fat{\beta} is \fat{6.5} for the constant-\fat{Q} model and \fat{8} for the linear model (constant \fat{\Delta t_j}). Thus, for a large \fat{\bar{a}} or a strong anti-Lagrange tendency, co-orbital planetary systems may survive for the entire life-time of the star in the main sequence.

\begin{figure}[h]
	\centering
	\includegraphics[width=0.9\linewidth]{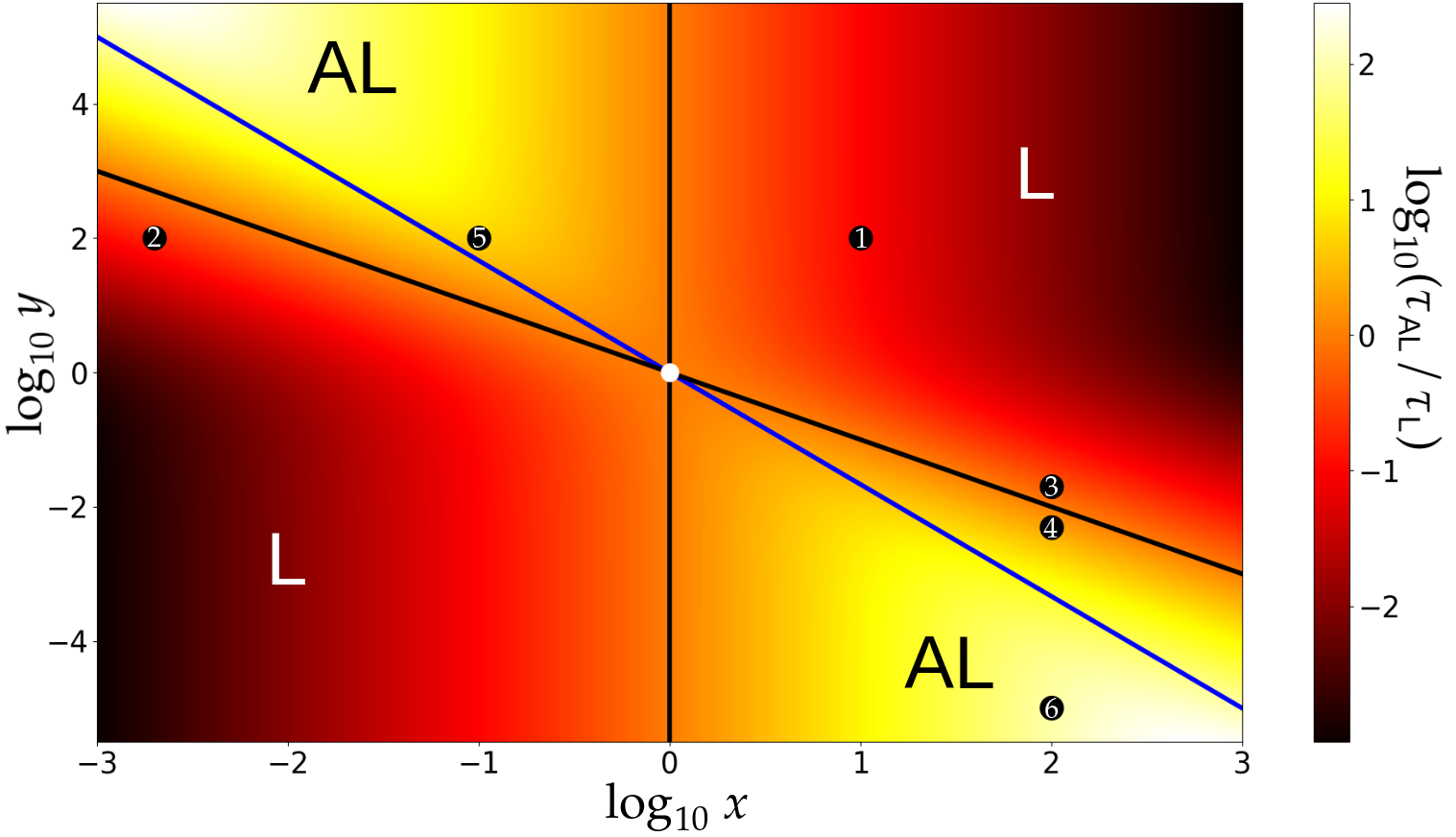}
	\caption[]{Decimal logarithm of \fat{\tau_{\text{AL}}/\tau_{\text{L}}} in colorscale. The black straight lines are the locations of the points where \fat{\tau_{\text{AL}}=\tau_{\text{L}}}. The blue line is the path followed by a system of variable mass repartition. The black spots are the positions of the numerical simulations of section \ref{S4}. The white dot \fat{\left(x=1,y=1\right)} is where \citet{RodriguezGiupponeMichtchenko2013} performed all of their simulations. It complies with \fat{\tau_{\text{AL}}=\tau_{\text{L}}}, explaining why they did not see any hierarchy between the eccentric eigen-modes. Systems in the red-black regions settle into Lagrange, have a short life expectancy and are eccentric when old, while systems in the yellow-white regions settle into anti-Lagrange, have a long life expectancy and are circular when old.}\label{tauALtauL}
\end{figure}

Interestingly, the ratio between the two eccentric damping timescales depends only on the ratio between the planetary masses and the ratio between the dissipation rates inside the planets. That is, denoting
\begin{equation}\label{xy}
x=\frac{m_1}{m_2}\place\text{and}\place y=\frac{q_2Q_1}{q_1Q_2}=\frac{q_2\eta\Delta t_2}{q_1\eta\Delta t_1}=\frac{\kappa_{2,2}\qoppa_2^5\Delta t_2}{\kappa_{2,1}\qoppa_1^5\Delta t_1},
\end{equation}
we have
\begin{equation}
\frac{\tau_{\text{AL}}}{\tau_{\text{L}}}=\frac{x\left(1+y\right)}{1+yx^2}.
\end{equation}
The equality between the eccentric damping timescales (\fat{\tau_{\text{AL}}=\tau_{\text{L}}}) occurs at \fat{x=1} and \fat{xy=1}, plotted by black lines in figure \ref{tauALtauL}, where we also show the ratio \fat{\tau_{\text{AL}}/\tau_{\text{L}}} as a function of \fat{x} and \fat{y}. We clearly observe two regions, corresponding to Lagrange-like and anti-Lagrange-like systems.

Let us assume that the quality factor \fat{Q_j} is mass independent. We then deduce from \fat{\qoppa_j\propto m_j^{1/3}} and equation (\ref{xy}) that
\begin{equation}
y\propto x^{-5/3}.
\end{equation}
The blue line in figure \ref{tauALtauL} plots \fat{y=x^{-5/3}}. It shows the path followed by a system with two initially identical planets (white spot) when we change the mass repartition between them. We conclude that for a given sum of the planetary masses \fat{\varepsilon} and sum of the dissipation rates  \fat{q_1/Q_1+q_2/Q_2}, the expectancy of life of the system is at its shortest for \fat{m_1=m_2=\varepsilon m_0/2} (orange area) and it tends towards infinity if either \fat{m_1} or \fat{m_2} tends towards \fat{\varepsilon m_0} (yellow/white area).
\subsection{Application to the detection of co-orbital exoplanets}\label{tool}

We can use the results from section~\ref{tidal_impact} to ascertain if an already discovered exoplanet may have an undetected co-orbital companion. For any co-orbital system, we denote \fat{\tau_{\text{hs}}} and \fat{\tau_{\text{dest}}} the time needed to reach the horseshoe-shaped orbits, and the time needed for close encounters to disrupt the co-orbital resonance, respectively. In the rest of this section, we use the observational parameters to estimate \fat{\tau_{\text{dest}}} and then discard cases such that \fat{\tau_{\text{dest}}} is too small.

The libration amplitude $\Phi=\max\left(\xi-60^{\circ}\right)$ is defined as the angular distance to \fat{L_4}. If \fat{\Phi_0} is the initial libration amplitude of the system and if \fat{\Phi} is a small libration amplitude greater than \fat{\Phi_0}, then, by definition of \fat{\tau_{\text{lib}}}, the time \fat{\tau_{\Phi}} needed to reach the libration amplitude \fat{\Phi} is given by
\begin{equation}\label{tau_phi}
\frac{\tau_{\Phi}}{\tau_{\text{lib}}}=\ln\left(\frac{\Phi}{\Phi_0}\right).
\end{equation}
Equation (\ref{tau_phi}) a priori cannot be used to predict the time \fat{\tau_{\text{hs}}}, since the separatrix \fat{L_3} is far from the fixed point \fat{L_4}. Nevertheless, by performing numerical simulations of equations (\ref{equation_case1}) with arbitrary parameters and initial conditions, we verify that the expression
\begin{equation}\label{tau_hs}
\frac{\tau_{\text{hs}}}{\tau_{\text{lib}}}\approx\ln\left(\frac{60}{\Phi_0}\right)\approx 4.1-\ln\left(\Phi_0\right),
\end{equation}
is always a good approximation as long \fat{\Phi_0\leq 15^{\circ}}, where \fat{\Phi_0} is in degrees.
Although \fat{\tau_{\text{hs}}/\tau_{\text{lib}}} only depends on \fat{\Phi_0}, we expect that \fat{\tau_{\text{dest}}/\tau_{\text{lib}}} also depends on \fat{\varepsilon}, since this parameter controls the maximum libration amplitude before the system becomes unstable. The smaller \fat{\eps} is, the larger the maximum libration amplitude is. Since \fat{\varepsilon=3\times10^{-4}} is the highest value allowing horseshoe-shaped orbits \citep{LeRoCo2015}, the equality \fat{\tau_{\text{dest}}=\tau_{\text{hs}}} occurs at this value. If we fix \fat{\Phi_0=10^{\circ}}, we have
\begin{equation}
1/2\,\tau_{\text{hs}}\leq\tau_{\text{dest}}\leq 2\,\tau_{\text{hs}}\;\Leftrightarrow\; 10^{-9}\lesssim\eps\lesssim 0.005,
\end{equation}
and so, \fat{\tau_{\text{hs}}} and \fat{\tau_{\text{dest}}} do not differ by more than a factor 2 for a wide range of \fat{\eps}. We thus consider that \fat{\tau_{\text{dest}}\approx\tau_{\text{hs}}} in this range and equations (\ref{tau_lib}) and (\ref{tau_hs}) can be used to predict \fat{\tau_{\text{dest}}}.
\begin{figure}[h]
	\centering
	\includegraphics[width=0.9\linewidth]{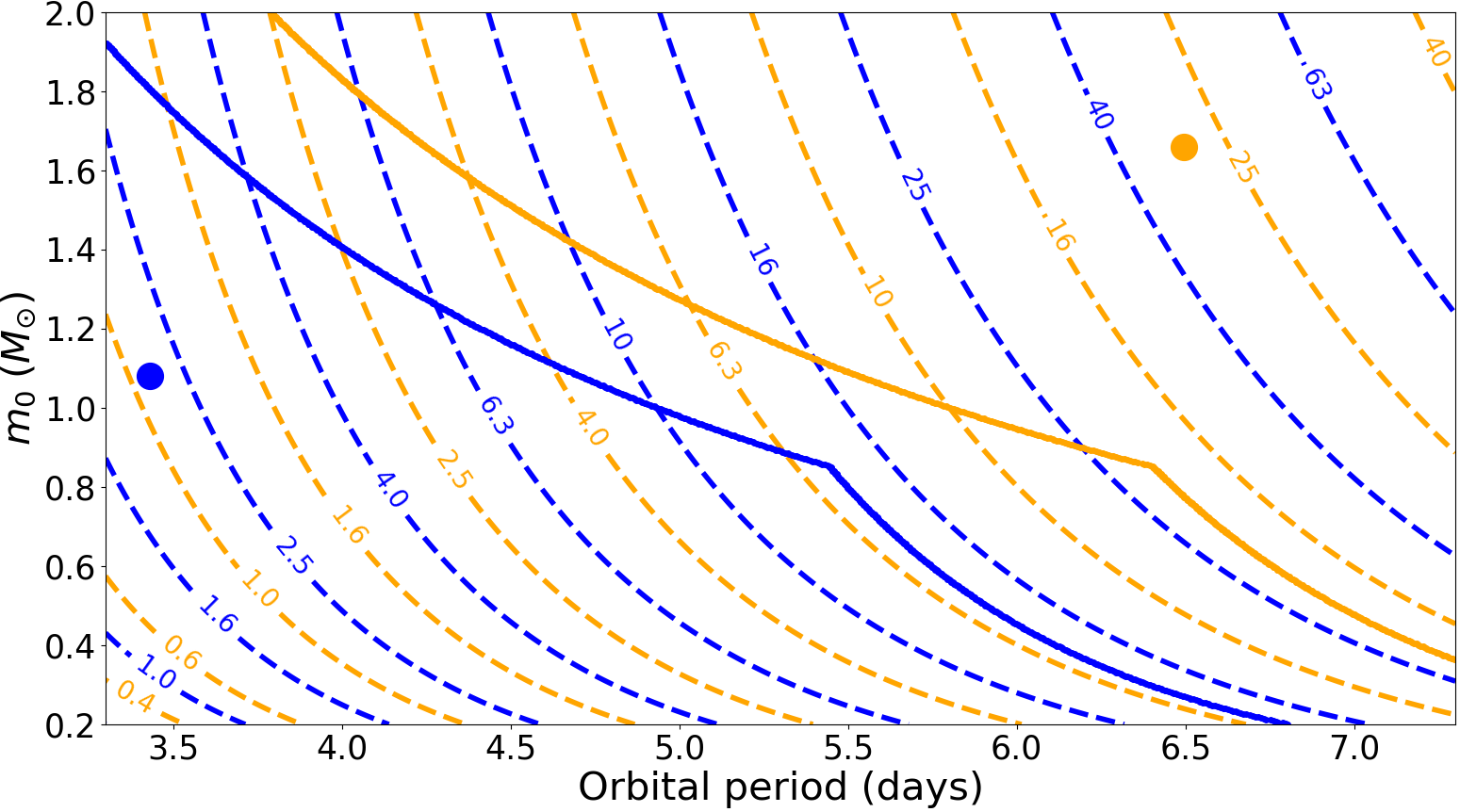}
	\caption[]{Disruption time (Gyr) of an hypothetical co-orbital system as a function of the orbital period of the observed planet and of the mass of the host star. The couple Saturn+Earth is plotted in orange while the couple Earth+Earth is plotted in blue. The solid lines plot the minimum of the main sequence duration and of the age of the universe for both couples. Systems below these lines may have been already destroyed at the time of observation, but systems above outlive either their host star or the age of the universe. The blue and orange dots correspond to HD 158259 c and HD 102956 b, respectively. Note that, for a given orbital period, co-orbital systems live longer around a massive host star because they have a larger semi-major axis.}\label{tau_dest}
\end{figure}

Figure \ref{tau_dest} shows the disruption times of a Saturn-mass and an Earth-mass co-orbital system (in orange), and of a system of two Earth-mass (in blue), with \fat{\Phi_0=10^{\circ}}. The exact parameters adopted for these systems are listed in Table \ref{bodies}. The solid line plots \fat{\min\left(\tau_{\text{ms}},\tau_{\text{u}}\right)} where \fat{\tau_{\text{ms}}=10^{10}\,\text{yr}\,\left(m_0/M_{\odot}\right)^{-2}} is the duration of the main-sequence of the host star and \fat{\tau_{\text{u}}=13.77}~Gyr is the age of the universe. Figure \ref{tau_dest} depends little on the choice of the gas giant and on the choice of the rocky body, in the sense that it would not be drastically different if other planets had been chosen. Therefore, it tells us if an already detected gas giant may have a companion (orange lines), or if an already detected rocky planet may have a companion (blue lines). The detected exoplanet is located in figure \ref{tau_dest} using its orbital period and the mass of the host star. If it is below its associated solid line (orange for a gas giant, blue for a rocky planet), then its companion was already ejected, if it ever existed. On the contrary, if it is above this line, the subsequent pair of co-orbital outlives either the host star or the age of the universe and it is worth looking for the companion.
As an example, it is very unlikely to find a co-orbital companion for the rocky planet HD 158259 c \citep{Hara2020}, plotted with a blue dot in figure \ref{tau_dest}, but we cannot rule out that the gas giant HD 102956 b \citep{Luhn2019}, plotted with a orange dot in figure \ref{tau_dest}, has a co-orbital companion.
\begin{table}
\begin{center}
\begin{tabular}{|r|r|r|r|r|}
\hline
&mass \fat{\left(M_{\oplus}\right)}&specific mass (kg\fat{/}m\fat{^3})&\fat{\kappa_2}&\fat{Q}\\
\hline
Earth&1&5515&0.302&280\\
\hline
Saturn&95.15&687.3&0.39&2450\\
\hline
\end{tabular}
\caption{Parameters of the bodies used to plot figure \ref{tau_dest}. \citep{Lainey2016}}\label{bodies}
\end{center}
\end{table}

\section{Numerical simulations}\label{S4}
In this section, in order to verify results of section \ref{sec:tide} and expand them to the whole space phase, we perform some numerical simulations of planetary systems representative of the different dynamical regimes of co-orbital planets undergoing tidal interactions with the star, such as those described in section \ref{sec:tide}. More precisely, we display the evolution of the six systems that correspond to the black dots on figure \ref{tauALtauL}. For all of them we choose $\varepsilon=2\times 10^{-4}$, $q_1/Q_1+q_2/Q_2=4\times 10^{-13}$, $m_0=M_{\odot}$, $\rho_1=500\,\text{kg}/\text{m}^3$, $\rho_2=2000\,\text{kg}/\text{m}^3$, $\alpha_1=\alpha_2=0.33$, $e_{2,0}=2e_{1,0}=0.04$, $\varpi_{1,0}=\varpi_{2,0}=0$, $a_{1,0}=a_{2,0}=\bar{a}=0.02$ AU, $\vartheta_{1,0}=\vartheta_{2,0}=0$ and $\xi_0=62^{\circ}$, that is, the systems are initially \fat{2^{\circ}} away from \fat{L_4}. Since the total planetary masses \fat{\varepsilon} and the total dissipation rate \fat{q_1/Q_1+q_2/Q_2} are the same for all systems, their positions can all be plotted in figure \ref{tauALtauL} and their tidal timescales are entirely determined by the values of the mass ratio \fat{x} and the dissipation rate ratio \fat{y} (see Eq. (\ref{xy})), which are the only variable parameters (see table \ref{table_xy}). The choice for the initial eccentricities and longitude of the perihelions guarantees that the systems are not initially collinear to the Lagrange or anti-Lagrange configuration, and their evolution allows us to know if they are Lagrange-like or anti-Lagrange-like.
\begin{figure}[h]
	\centering
	\includegraphics[width=0.9\linewidth]{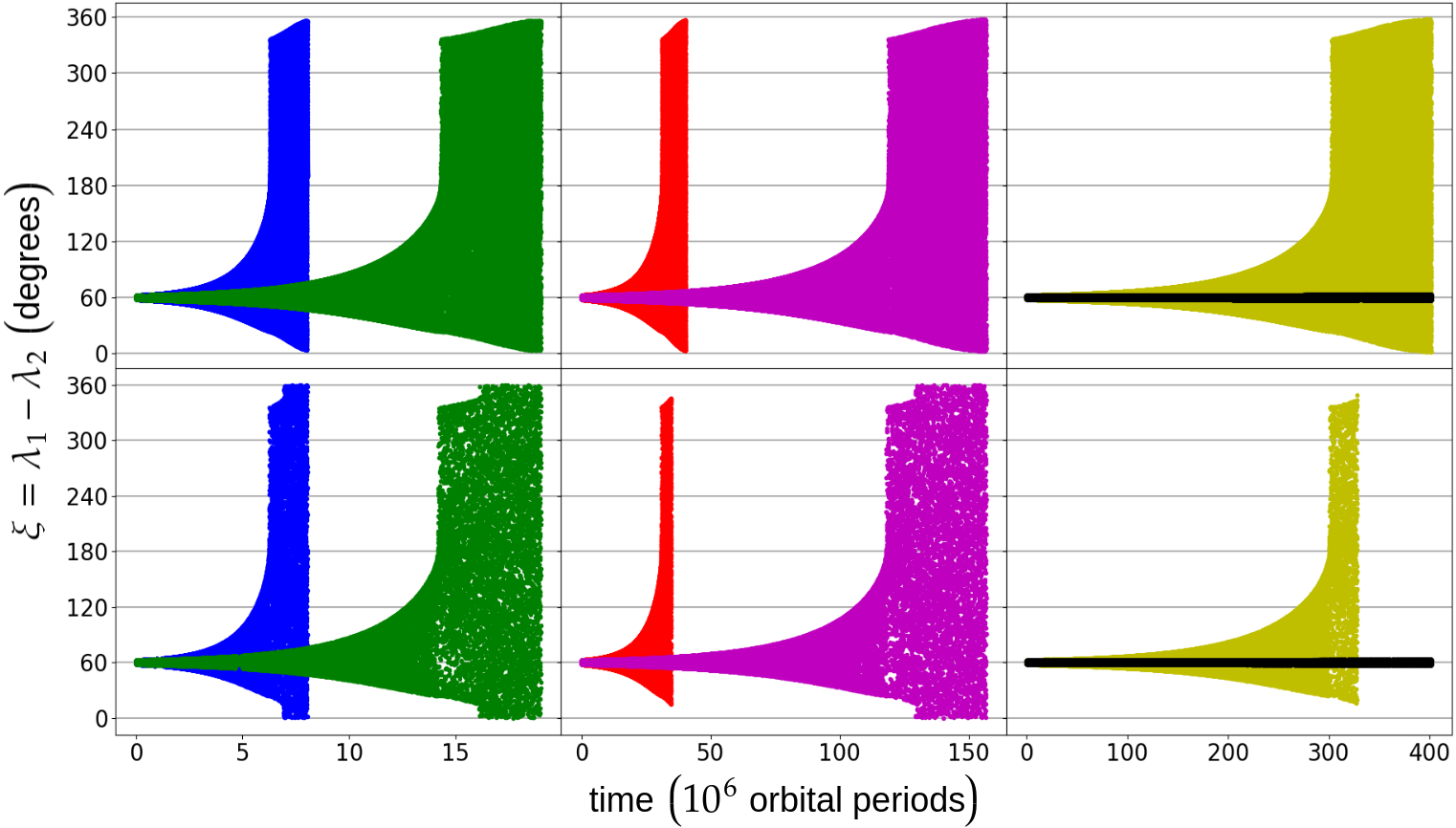}
	\caption[]{Value of the libration angle \fat{\xi=\lambda_1-\lambda_2} against time. The switching from tadpole to horseshoe orbits (crossing of the separatrix emanating from \fat{L_3}) occurs when the libration amplitude suddenly increases, giving this boot-shape to the plots. As expected from figure \ref{tauALtauL} and table \ref{table_xy}, systems 4, 5 and 6 live longer than systems 1, 2 and 3.}\label{xi}
\end{figure}

\begin{figure}[h]
	\centering
	\includegraphics[width=0.9\linewidth]{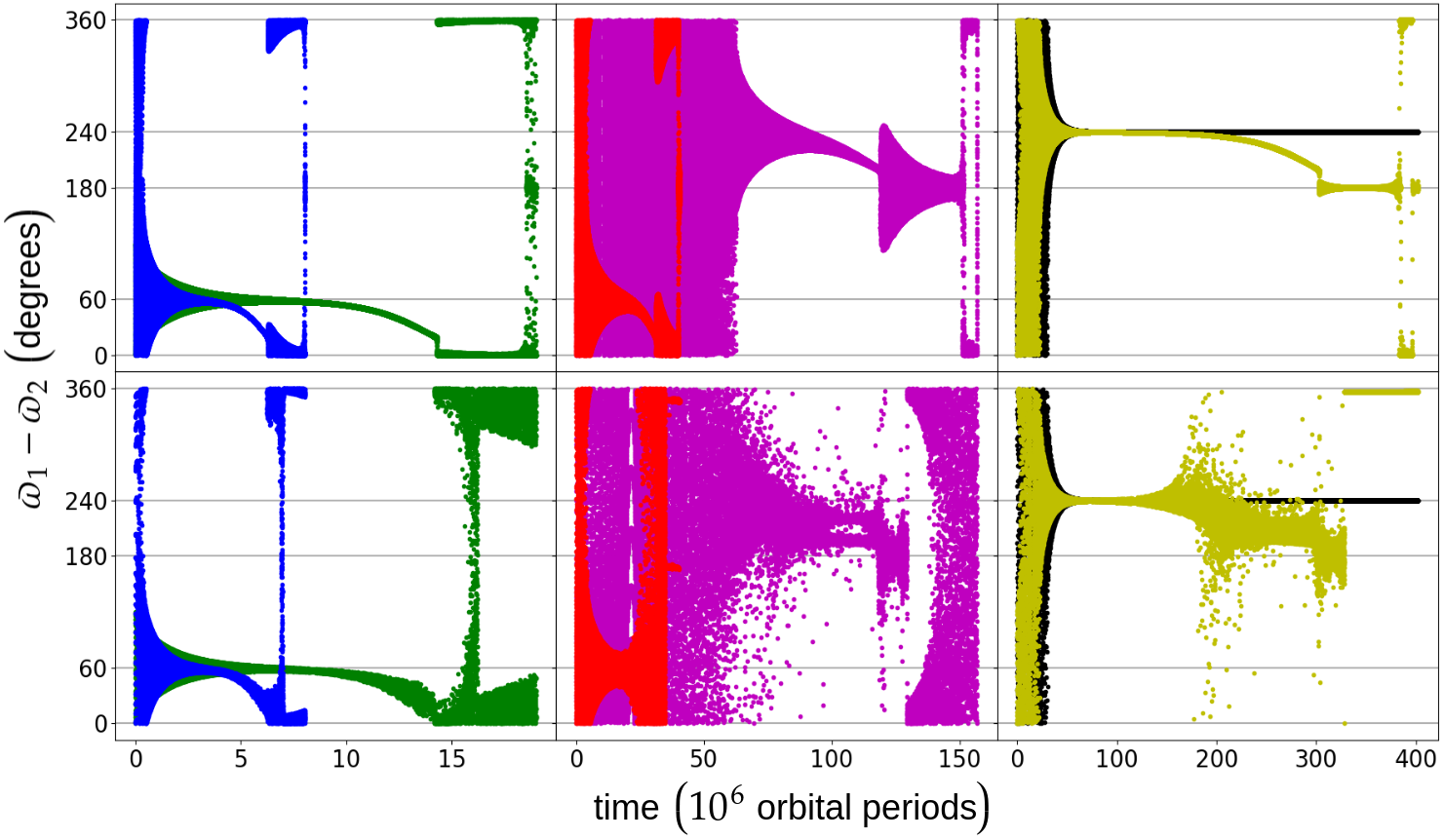}
	\caption[]{Value of \fat{\varpi_1-\varpi_2} against time. Blue, green and red systems settle into Lagrange and have a short life while purple, yellow and black systems settle into anti-Lagrange and live long. This is in agreement with what we deduce from figure \ref{tauALtauL} and table \ref{table_xy}.}\label{vp}
\end{figure}

\begin{figure}[h]
	\centering
	\includegraphics[width=0.9\linewidth]{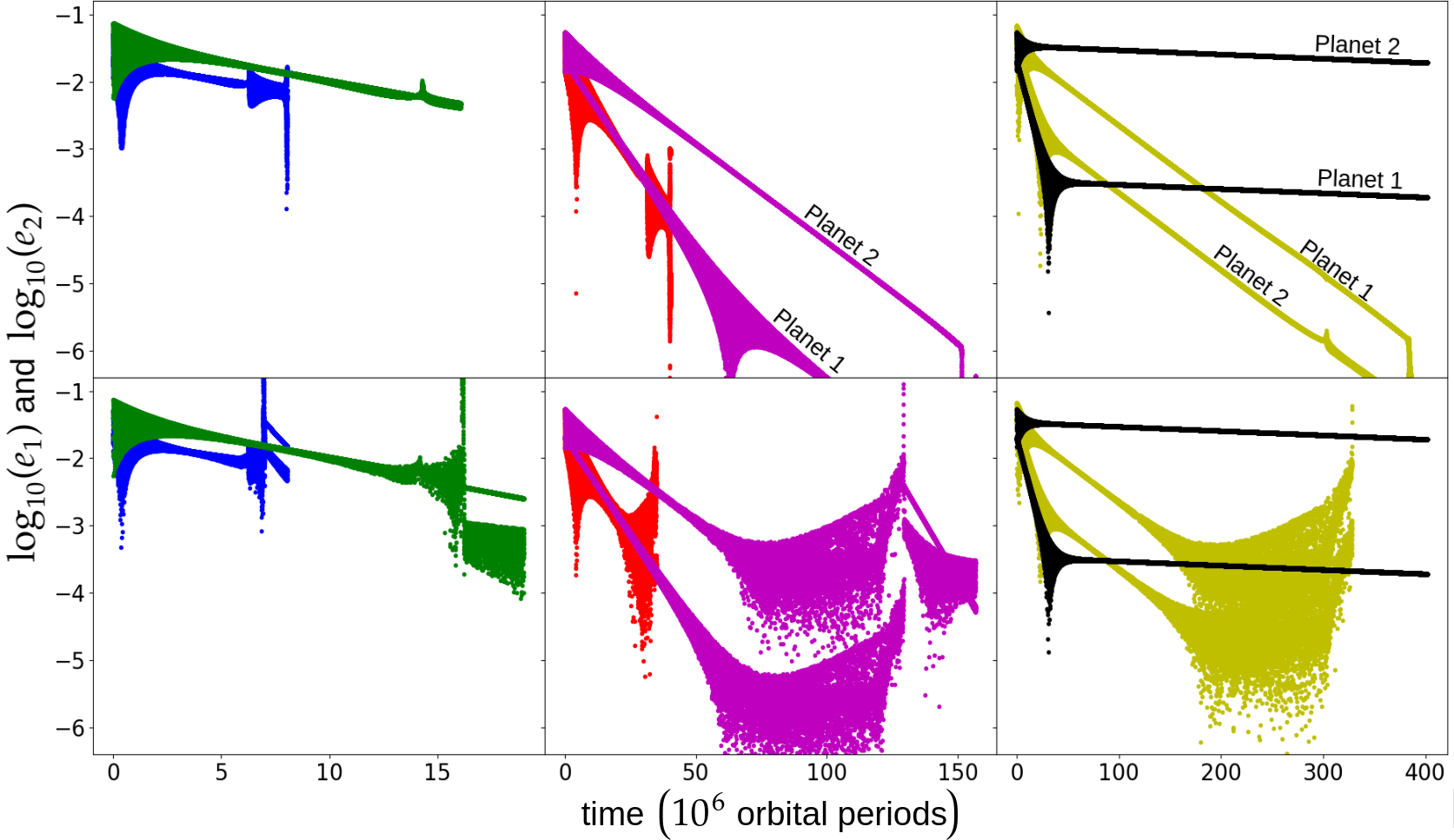}
	\caption[]{Eccentricities of both planets against time. For a same system, \fat{e_1} and \fat{e_2} are plotted with the same color. Anti-Lagrange-like systems comply with \fat{e_2=xe_1} and both plots are easily distinguished while for Lagrange-like system, such that \fat{e_1=e_2}, both plots are almost overlaid. This is in agreement with what we deduce from figure \ref{tauALtauL} and table \ref{table_xy}.}\label{e}
\end{figure}

The timescales of the six systems, expressed in number of orbital periods and deduced from section \ref{tidal_times}, are given in table \ref{table_xy}. \fat{\tau_{\text{lib}}} corresponds to the characteristic timescale of libration amplitude excitation given by equations (\ref{timescales}), while \fat{\tau_{\text{lib}}^{\text{num}}} is directly computed from the numerical value of \fat{\mathcal{Z}_0+\mathcal{Z}_1} (given in appendix \ref{append_linear}), without considering any first-order expansion.
\begin{table}[h]
\begin{center}
\begin{tabular}{|r|r|r|r|r|r|r|r|}
\hline
\#&color&\fat{x}&\fat{y}&\fat{\tau_{\text{lib}}}&\fat{\tau_{\text{lib}}^{\text{num}}}&\fat{\tau_{\text{AL}}}&\fat{\tau_{\text{L}}}\\
\hline
1&blue&10&100&1\,785\,893&1\,845\,021&765\,382&7\,578\,807\\
\hline
2&green&1\fat{/}500&100&3\,570\,716&4\,198\,710&1\,530\,306&7\,578\,807\\
\hline
3&red&100&1\fat{/}50&8\,973\,910&9\,161\,859&3\,845\,961&7\,578\,807\\
\hline
4&purple&100&1\fat{/}200&34\,847\,651&34\,893\,952&14\,934\,707&7\,578\,807\\
\hline
5&yellow&1\fat{/}10&100&89\,303\,607&89\,304\,263&38\,272\,974&7\,578\,807\\
\hline
6&black&100&\fat{10^{-5}}&1\,607\,641\,764&1\,607\,642\,323&688\,989\,327&7\,578\,807\\
\hline
\end{tabular}
\caption{Values of \fat{x}, \fat{y} and of the corresponding timescales of the systems. The timescales are expressed in number of orbital periods.}\label{table_xy}
\end{center}
\end{table}
For some systems, especially system 2, there is a slight difference between \fat{\tau_{\text{lib}}^{\text{num}}} and \fat{\tau_{\text{lib}}}. For these systems, the values of \fat{d_1} and \fat{d_2} in the matrix \fat{\mathcal{Z}_1} are not so small with respect to \fat{1}, leading to a value for \fat{\zeta} also not so small with respect to 1 (see appendix \ref{append_diago}). A smaller value for the sum of the dissipation rates \fat{q_1/Q_1+q_2/Q_2} would provide a better agreement, but it also leads to much longer simulations. 
On the other hand, there is no disagreement between the analytical values of \fat{\tau_{\text{AL}}} and \fat{\tau_{\text{L}}} in section \ref{tidal_times} and their numerical counterparts, since the eccentricities are uncoupled from the rest of the variables in the linearized system.
\begin{table}[h]
	\begin{center}
		\begin{tabular}{|l|r|r|r|r|}
			\hline
			\#&\fat{\tau_{\text{hs}}^{\ref{equation_case1}}} \fat{\left(10^6\;\text{periods}\right)}&\fat{\tau_{\text{hs}}^{\ref{nbody_direct}}} \fat{\left(10^6\;\text{periods}\right)}&relative error&\fat{\tau_{\text{dest}}} \fat{\left(10^6\;\text{periods}\right)}\\
			\hline
			1&6.28560&6.24269&0.683 \%&6.92019\\
			\hline
			2&14.3061&14.1782&0.894 \%&16.1457\\
			\hline
			3&31.2717&31.0543&0.695 \%&34.8405\\
			\hline
			4&118.975&118.150&0.693 \%&129.357\\
			\hline
			5&302.789&300.438&0.777 \%&328.044\\
			\hline
			6&around 5\,466 (see (\ref{tau_hs}))&&&around 6\,010\\
			\hline
		\end{tabular}
		\caption{Times to reach horseshoe-shaped orbits and until destruction.}\label{table_tau_hs}
	\end{center}
\end{table}
Each system is numerically integrated using two different sets of equations. In the first set, we use the secular equations (\ref{equation_case1}) derived in this paper. In the second set, we use a \fat{n}-body direct model (see Appendix \ref{append_nbody_direct}, Eq. (\ref{nbody_direct})). The results are displayed in figures \ref{xi}, \ref{vp} and \ref{e}. The top panels correspond to the set of equations (\ref{equation_case1}) while the bottom panels correspond to the set (\ref{nbody_direct}). All systems except system~6 are integrated long enough for the co-orbital configuration to be destroyed.

We observe there is always a very good agreement between equations (\ref{equation_case1}) and equations (\ref{nbody_direct}), except when the libration amplitude is near \fat{360^{\circ}}. Indeed, as the libration amplitude increases, close encounters between the planets mean that planet-planet interactions are no longer perturbations of the Keplerian motion and our model is no longer valid. On the upper plot of figure \ref{xi}, the amplitude of \fat{\xi} tends towards \fat{2\pi} when time goes to infinity but never reaches it, while on the lower plot, there exist a finite time for which \fat{\xi} reaches \fat{2\pi}, meaning it is a circulating angle and the motion is not in a 1:1 resonance anymore. All simulations confirm that the co-orbital resonance is left.

We also note that for small eccentricities, averaged equations (\ref{equation_case1}) (plotted on top) differ significantly from the direct equations (\ref{nbody_direct}) (plotted at the bottom). We believe this is due to short period influences that were averaged in our model. Indeed, while the manifold \fat{\left(X_1=0,\,X_2=0\right)} is stable by the flow of the averaged Hamiltonian, it is not stable by the flow of the complete Hamiltonian. This may explain the differences between both plots at low eccentricities. This does not discredit though the theoretical results obtained in section \ref{sec:tide}.

In table \ref{table_tau_hs}, we give for reference  the times \fat{\tau_{\text{hs}}^{\ref{equation_case1}}} and \fat{\tau_{\text{hs}}^{\ref{nbody_direct}}} needed to reach horseshoe-shaped orbits as deduced from the simulations of the equations (\ref{equation_case1}) and the direct \fat{n}-body simulations (\ref{nbody_direct}), respectively, as well as the time \fat{\tau_{\text{dest}}} until destruction of the co-orbital configuration, deduced from equations (\ref{nbody_direct}).
The relative error on the time of crossing of the separatrix \fat{L_3} is consistently smaller than 1\%, showing the reliability of the model on the whole tadpole region for this choice of \fat{\varepsilon}. We have, for the first 5 systems, \fat{\tau_{\text{hs}}/\tau_{\text{lib}}^{\text{num}}=3.4}, which is consistent with equation (\ref{tau_hs}).

As expected from the values of \fat{\tau_{\text{AL}}} and \fat{\tau_{\text{L}}} given in table~\ref{table_xy}, the three shortest simulations correspond to Lagrange-like systems, while the three longest correspond to anti-Lagrange like systems. This is particularly clear in Figure \ref{vp}, which displays the value of \fat{\varpi_1-\varpi_2}. For the short-lived systems (blue, green and red), the difference of the longitude of perihelion first settles around \fat{60^{\circ}=\pi/3} at low libration amplitude, before it moves to \fat{0^\circ}, when in horseshoe orbit. For the long-lived systems (yellow, black and purple), the difference of the longitude of perihelion first settles around \fat{240^{\circ}=4\pi/3} at low libration amplitude, before it moves to \fat{180^{\circ}=\pi}, when in horseshoe orbit. This is in complete accordance with Eqs. (\ref{pisur3}), (\ref{4pisur3}), (\ref{zero}) and (\ref{pi}). 

Figure \ref{e} also confirms the Lagrange-like behaviour of short-lived systems, since \fat{e_1 = e_2} for these systems, while \fat{m_1 e_1 = m_2 e_2} for long-lived simulations, characteristic of their anti-Lagrange-like behaviour. As also expected from the theoretical results, Lagrange-like system are still eccentric when they are old (see for example the green and blue plots), while anti-Lagrange-like system are circular when they are at the end of their life (see the yellow plot). 
\begin{figure}[h]
	\centering
	\includegraphics[width=0.8\linewidth]{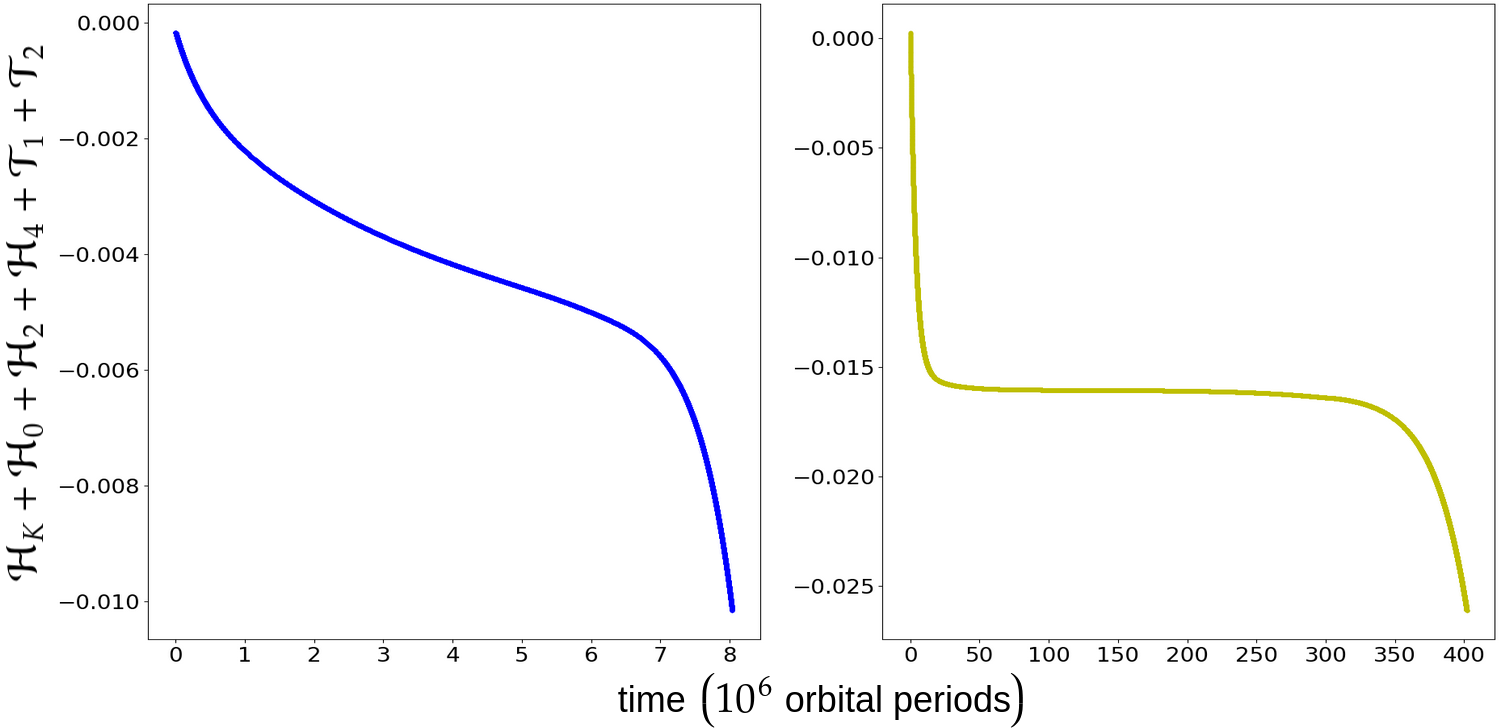}
	\caption[]{Total normalized energy of system 1 (left) and system 5 (right). The very anti-Lagrange-like system 5 features a pronounced plateau in its energy decrease and lives long while the very Lagrange-like system 1 does not show any marked plateau and has a short life.}\label{Energy}
\end{figure}

In Figure \ref{Energy}, we show the total energy \fat{\Ham_K+\Ham_0+\Ham_2+\Ham_4+\mathcal{T}_1+\mathcal{T}_2} of system~1 and system~5. 
The energy of the anti-Lagrange-like system~5 (yellow) features two steep decrease and a broad plateau. At the beginning of the simulation, the eccentricities of the planets are still significant, meaning that \fat{\dot{w}_j}, the time derivative of their true longitude, is not constant. This prevents a solid rotation around the star, although the equilibrium rotation is already reached (small \fat{\tau_{\text{rot}}^j}), and ensures dissipation: it is the first steep decrease. Then, the eccentricities are almost damped while the libration amplitude did not significantly increase yet. The motion around the star is almost a solid rotation and very few energy is dissipated into heat: it is the plateau. Finally, the high libration amplitude of \fat{\xi} reached in late tadpole and horseshoe-shaped orbits ensures the non-constancy of \fat{\dot{w}_j} and energy dissipation: it is the second steep decrease. 
For almost restricted systems, very anti-Lagrange-like, the amplitude of libration has to be very high for the least massive planet to significantly perturb the solid rotation of the most massive one. Thus, these systems have very broad plateau and a very high life expectancy. On the other hand, the energy of the Lagrange-like system~1 (blue), do not feature such a marked plateau, because the eccentricities are always nonzero even at high libration amplitude, and hence its life is far shorter. 

This section shows that equations (\ref{equation_case1}) allow to correctly predict the tidal evolution of co-orbital systems.
One advantage of using these equations instead of the direct equations (\ref{nbody_direct}) is that they lead to less-oscillating curves. Another advantage is that they are quicker to run (by a factor \fat{1/\nu}), since the average over the fast dynamics allows to take a larger time-step. With \fat{\varepsilon=2\,10^{-4}}, equations (\ref{equation_case1}) are \fat{1/\nu=27.2} times faster to run than the direct ones with equivalent CPU performances. Finally, the average equations (\ref{equation_case1}) also allow us to better understand and predict the dynamical behaviour of co-orbital systems.

\section{Conclusion}\label{conclusion}
In this paper, we have studied the tidal evolution of co-orbital planetary systems. We have shown that these systems are always unstable, regardless of the parameters and initial configuration. However, for well chosen parameters, the disruption timescale can be arbitrarily high if the planets orbit far enough from the central star and/or if the mass repartition between both planets is far enough from equal masses. Since the current detection methods of exoplanets (radial velocity and transit) favour the detection of close-in systems with roughly equal masses for the planets, this work gives a satisfactory explanation to why no co-orbital planets have been discovered so far, although planetary formation models predict their formation. In figure \ref{tau_dest}, we give a useful plot to predict if an already detected exoplanet may have a co-orbital companion. 

We provide analytic criteria to determine the evolution timescales of a co-orbital system, which allow us to predict if a system is eccentric or circular when it is on the verge of being destroyed and the eccentric proper-mode it has settled into. We also show the difference between the two proper-mode according to whether the system is in the tadpole or horseshoe region. We show that the tidal evolution is essentially dominated by three characteristics timescales. \fat{\tau_{\text{L}}} and \fat{\tau_{\text{AL}}} are responsible for damping the proper-modes of the eccentricity and \fat{\tau_{\text{lib}}} for pumping the libration amplitude of the resonant angle. The latter is also responsible for the destruction of the co-orbital configuration, and it provides us with an estimate of its life time.

We obtain a complete system of averaged equations describing the evolution of the system which is consistent with the conservation of the total angular momentum at any order in eccentricity and yields no significant difference with respect to the \fat{n}-body direct equations when compared by numerical integrations. This shows that the averaging process over the orbital period, as well as all approximations made are correct. Therefore, this system of equations allows us to perform fast numerical integrations with a larger time-step.

In the present work we did not consider tides raised on the star (see Eq. (\ref{neglect_star})). We did not consider the general relativity and the rotational oblateness of the bodies either, since these conservative effects only slightly modify the imaginary parts of the eigenvalues of the linearized system (Eq.\,(\ref{eigenvalues_circular})), which only results in small changes in the fundamental frequencies \fat{\nu,\,g_1\,\text{and}\,g_2}, and does not impact the timescales of the co-orbital system.
The quasi-satellite configuration \citep{PousseRobutelVienne2017}, which is a peculiar type of 1:1 mean motion resonance that does not exist at zero eccentricity, was also not studied by this work, because our model is singular when the planets have the same mean longitude. However, since tidal effects damp the eccentricities of the planets, we believe that this type of orbit is even more unstable than the Lagrange and anti-Lagrange configurations. This was numerically confirmed by \cite{RodriguezGiupponeMichtchenko2013} in the case of two identical planets.

\bibliographystyle{apalike}
\bibliography{\bibpath biblio.bib}

\begin{thebibliography}{}

\bibitem[{Adams} and {Bloch}, 2015]{Adams_Bloch_2015}
{Adams}, F.~C. and {Bloch}, A.~M. (2015).
\newblock {On the stability of extrasolar planetary systems and other closely
  orbiting pairs}.
\newblock {\em Monthly Notices of the Royal Astronomical Society},
  446:3676--3686.

\bibitem[{Beaug{\'e}} et~al., 2007]{Beauge_etal_2007}
{Beaug{\'e}}, C., {S{\'a}ndor}, Z., {{\'E}rdi}, B., and {S{\"u}li}, {\'A}.
  (2007).
\newblock {Co-orbital terrestrial planets in exoplanetary systems: a formation
  scenario}.
\newblock {\em Astron. Astrophys.}, 463:359--367.

\bibitem[{Claret} and {Cunha}, 1997]{Claret_Cunha_1997}
{Claret}, A. and {Cunha}, N.~C.~S. (1997).
\newblock {Circularization and synchronization times in Main-Sequence of
  detached eclipsing binaries II. Using the formalisms by Zahn.}
\newblock {\em Astronomy and Astrophysics}, 318:187--197.

\bibitem[{Correia} et~al., 2014]{Correia_etal_2014}
{Correia}, A.~C.~M., {Bou{\'e}}, G., {Laskar}, J., and {Rodr{\'{\i}}guez}, A.
  (2014).
\newblock {Deformation and tidal evolution of close-in planets and satellites
  using a Maxwell viscoelastic rheology}.
\newblock {\em Astronomy and Astrophysics}, 571:A50.

\bibitem[{Correia} et~al., 2020]{Correia_etal_2020}
{Correia}, A.~C.~M., {Bourrier}, V., and {Delisle}, J.~B. (2020).
\newblock {Why do warm Neptunes present nonzero eccentricity?}
\newblock {\em Astronomy and Astrophysics}, 635:A37.

\bibitem[{Correia} and {Laskar}, 2004]{Correia_Laskar_2004}
{Correia}, A.~C.~M. and {Laskar}, J. (2004).
\newblock {Mercury's capture into the 3/2 spin-orbit resonance as a result of
  its chaotic dynamics}.
\newblock {\em Nature}, 429:848--850.

\bibitem[{Correia} and {Laskar}, 2010]{Correia_Laskar_2010B}
{Correia}, A.~C.~M. and {Laskar}, J. (2010).
\newblock {\em {Tidal Evolution of Exoplanets}}, pages 239--266.
\newblock University of Arizona Press, Tucson.

\bibitem[{Cresswell} and {Nelson}, 2008]{CreNe2008}
{Cresswell}, P. and {Nelson}, R.~P. (2008).
\newblock {Three-dimensional simulations of multiple protoplanets embedded in a
  protostellar disc}.
\newblock {\em Astron. Astrophys.}, 482:677--690.

\bibitem[{Danby}, 1964]{Dan1964}
{Danby}, J.~M.~A. (1964).
\newblock Stability of the triangular points in the elliptic restricted problem
  of three bodies.
\newblock {\em Astron. Astrophys.}, 69:165.

\bibitem[{Darwin}, 1880]{Darwin_1880}
{Darwin}, G.~H. (1880).
\newblock {On the secular change in the elements of a satellite revolving
  around a tidally distorted planet}.
\newblock {\em Philos. Trans. R. Soc. London}, 171:713--891.

\bibitem[{Efroimsky}, 2012]{Efroimsky_2012}
{Efroimsky}, M. (2012).
\newblock {Bodily tides near spin-orbit resonances}.
\newblock {\em Celestial Mechanics and Dynamical Astronomy}, 112:283--330.

\bibitem[Euler, 1764]{Euler1764}
Euler, L. (1764).
\newblock Considerationes de motu corporum coelestium.
\newblock {\em Novi commentarii academiae scientiarum Petropolitanae. Berlin
  acad.}, 10:544--558.

\bibitem[{Ford} and {Gaudi}, 2006]{FoGa2006}
{Ford}, E.~B. and {Gaudi}, B.~S. (2006).
\newblock {Observational Constraints on Trojans of Transiting Extrasolar
  Planets}.
\newblock {\em Astron. Astrophys.}, 652:L137--L140.

\bibitem[{Ford} and {Holman}, 2007]{FoHo2007}
{Ford}, E.~B. and {Holman}, M.~J. (2007).
\newblock {Using Transit Timing Observations to Search for Trojans of
  Transiting Extrasolar Planets}.
\newblock {\em Astron. Astrophys.}, 664(1):L51--L54.

\bibitem[{Gascheau}, 1843]{Ga1843}
{Gascheau}, G. (1843).
\newblock Examen d'une classe d'{\'e}quations diff{\'e}rentielles et
  application {\`a} un cas particulier du probl{\`e}me des trois corps.
\newblock {\em C. R. Acad. Sci. Paris}, 16(7):393--394.

\bibitem[Giuppone et~al., 2010]{Guippone2010}
Giuppone, C.~A., Beaug{\'e}, C., Michtchenko, T.~A., and {Ferraz-Mello}, S.
  (2010).
\newblock Dynamics of two planets in co-orbital motion.
\newblock {\em Monthly Notices of the Royal Astronomical Society},
  407:390--398.

\bibitem[{Giuppone} et~al., 2012]{GiuBe2012}
{Giuppone}, C.~A., {Benitez-Llambay}, P., , and {Beaug{\'e}}, C. (2012).
\newblock Origin and detectability of co-orbital planets from radial velocity
  data.
\newblock {\em MNRAS}, 421(1):356--368.

\bibitem[Hara et~al., 2020]{Hara2020}
Hara, N.~C., Bouchy, F., Stalport, M., Boisse, I., Rodrigues, J., Delisle,
  J.-B., Santerne, A., Henry, G.~W., Arnold, L., {Astudillo-Defru}, N.,
  Borgniet, S., Bonfils, X., Bourrier, V., Brugger, B., Courcol, B., Dalal, S.,
  Deleuil, M., Delfosse, X., Demangeon, O., D{\'i}az, R.~F., Dumusque, X.,
  Forveille, T., H{\'e}brard, G., Hobson, M.~J., Kiefer, F., Lopez, T., Mignon,
  L., Mousis, O., Moutou, C., Pepe, F., Rey, J., Santos, N.~C., S{\'e}gransan,
  D., Udry, S., and Wilson, P.~A. (2020).
\newblock The {{SOPHIE}} search for northern extrasolar planets. {{XVI}}.
  {{HD}} 158259: {{A}} compact planetary system in a near-3:2 mean motion
  resonance chain.
\newblock {\em Astronomy and Astrophysics}, 636:L6.

\bibitem[{Hippke} and {Angerhausen}, 2015]{HiAn2015}
{Hippke}, M. and {Angerhausen}, D. (2015).
\newblock {A Statistical Search for a Population of Exo-Trojans in the Kepler
  Data Set}.
\newblock {\em ApJ}, 811:1.

\bibitem[{Hut}, 1980]{Hut_1980}
{Hut}, P. (1980).
\newblock {Stability of tidal equilibrium}.
\newblock {\em Astronomy and Astrophysics}, 92:167--170.

\bibitem[{Janson}, 2013]{Janson2013Apj}
{Janson}, M. (2013).
\newblock {A Systematic Search for Trojan Planets in the Kepler Data}.
\newblock {\em APJ}, 774:156.

\bibitem[{Kaula}, 1964]{Kaula_1964}
{Kaula}, W.~M. (1964).
\newblock {Tidal dissipation by solid friction and the resulting orbital
  evolution}.
\newblock {\em Revs. Geophys.}, 2:661--685.

\bibitem[Lagrange, 1772]{Lagrange1772}
Lagrange (1772).
\newblock {\em \OE uvres compl{\`e}tes}.
\newblock Gouthier-Villars, Paris (1869).

\bibitem[Lainey, 2016]{Lainey2016}
Lainey, V. (2016).
\newblock Quantification of tidal parameters from {{Solar System}} data.
\newblock {\em Celestial Mechanics and Dynamical Astronomy}, 126:145--156.

\bibitem[Laskar et~al., 2012]{LaskarBoueCorreia2012}
Laskar, J., Bou{\'e}, G., and Correia, A. C.~M. (2012).
\newblock Tidal dissipation in multi-planet systems and constraints on orbit
  fitting.
\newblock {\em Astronomy and Astrophysics}, 538:A105.

\bibitem[Laskar and Robutel, 1995]{LaskarRobutel1995}
Laskar, J. and Robutel, P. (1995).
\newblock Stability of the {{Planetary Three}}-{{Body Problem}}. {{I}}.
  {{Expansion}} of the {{Planetary Hamiltonian}}.
\newblock {\em Celestial Mechanics and Dynamical Astronomy}, 62:193--217.

\bibitem[{Laughlin} and {Chambers}, 2002]{LauCha2002}
{Laughlin}, G. and {Chambers}, J.~E. (2002).
\newblock {Extrasolar Trojans: The Viability and Detectability of Planets in
  the 1:1 Resonance}.
\newblock {\em Astron. J.}, 124:592--600.

\bibitem[{Leleu} et~al., 2019]{LeCoAt2019}
{Leleu}, A., {Coleman}, G. A.~L., and {Ataiee}, S. (2019).
\newblock {Stability of the co-orbital resonance under dissipation. Application
  to its evolution in protoplanetary discs}.
\newblock {\em Astron. Astrophys.}, 631:A6.

\bibitem[{Leleu} et~al., 2015]{LeRoCo2015}
{Leleu}, A., {Robutel}, P., and {Correia}, A.~C.~M. (2015).
\newblock {Detectability of quasi-circular co-orbital planets. Application to
  the radial velocity technique}.
\newblock {\em Astron. Astrophys.}, 581:A128, 14pp.

\bibitem[Leleu et~al., 2018]{LeRoCo2018}
Leleu, A., Robutel, P., and Correia, A. C.~M. (2018).
\newblock On the coplanar eccentric non-restricted co-orbital dynamics.
\newblock {\em Celestial Mechanics and Dynamical Astronomy}, 130:24.

\bibitem[{Leleu} et~al., 2017]{LeRoCoLi2017}
{Leleu}, A., {Robutel}, P., {Correia}, A.~C.~M., and {Lillo-Box}, J. (2017).
\newblock Detection of co-orbital planets by combining transit and
  radial-velocity measurements.
\newblock {\em Astron. Astrophys.}, 599:L7.

\bibitem[{Lillo-Box} et~al., 2018a]{LiBaFiLeSaCoRoFa2018}
{Lillo-Box}, J., {Barrado}, D., {Figueira}, P., {Leleu}, A., {Santos}, N.~C.,
  {Correia}, A.~C.~M., {Robutel}, P., and {Faria}, J.~P. (2018a).
\newblock {The TROY project: Searching for co-orbital bodies to known planets.
  I. Project goals and first results from archival radial velocity}.
\newblock {\em Astron. Astrophys.}, 609:A96.

\bibitem[{Lillo-Box} et~al., 2018b]{LiLePaFietal2018}
{Lillo-Box}, J., {Leleu}, A., {Parviainen}, H., {Figueira}, P., {Mallonn}, M.,
  {Correia}, A.~C.~M., {Santos}, N.~C., {Robutel}, P., {Lendl}, M., {Boffin},
  H.~M.~J., {Faria}, J.~P., {Barrado}, D., and {Neal}, J. (2018b).
\newblock {The TROY project. II. Multi-technique constraints on exotrojans in
  nine planetary systems}.
\newblock {\em Astron. Astrophys.}, 618:A42.

\bibitem[Luhn et~al., 2019]{Luhn2019}
Luhn, J.~K., Bastien, F.~A., Wright, J.~T., Johnson, J.~A., Howard, A.~W., and
  Isaacson, H. (2019).
\newblock Retired {{A Stars}} and {{Their Companions}}. {{VIII}}. 15 {{New
  Planetary Signals}} around {{Subgiants}} and {{Transit Parameters}} for
  {{California Planet Search Planets}} with {{Subgiant Hosts}}.
\newblock {\em The Astronomical Journal}, 157:149.

\bibitem[{Lyra} et~al., 2009]{LyJo2009}
{Lyra}, W., {Johansen}, A., {Klahr}, H., and {Piskunov}, N. (2009).
\newblock {Standing on the shoulders of giants. Trojan Earths and vortex
  trapping in low mass self-gravitating protoplanetary disks of gas and
  solids}.
\newblock {\em Astron. Astrophys.}, 493:1125--1139.

\bibitem[{Madhusudhan} and {Winn}, 2009]{MaWi2009}
{Madhusudhan}, N. and {Winn}, J.~N. (2009).
\newblock {Empirical Constraints on Trojan Companions and Orbital
  Eccentricities in 25 Transiting Exoplanetary Systems}.
\newblock {\em APJ}, 693(1):784--793.

\bibitem[Mignard, 1979]{Mignard1979}
Mignard, F. (1979).
\newblock The evolution of the lunar orbit revisited. {{I}}.
\newblock {\em Moon and Planets}, 20:301--315.

\bibitem[Moeckel, 2017]{Moeckel2017}
Moeckel, R. (2017).
\newblock Minimal energy configurations of gravitationally interacting rigid
  bodies.
\newblock {\em Celestial Mechanics and Dynamical Astronomy}, 128:3--18.

\bibitem[{Munk} and {MacDonald}, 1960]{Munk_MacDonald_1960}
{Munk}, W.~H. and {MacDonald}, G.~J.~F. (1960).
\newblock {\em {The Rotation of the Earth; A Geophysical Discussion}}.
\newblock Cambridge University Press.

\bibitem[{Namouni}, 1999]{Namouni1999}
{Namouni}, F. (1999).
\newblock {Secular Interactions of Coorbiting Objects}.
\newblock {\em Icarus}, 137:293--314.

\bibitem[{Nauenberg}, 2002]{Nauenberg2002}
{Nauenberg}, M. (2002).
\newblock {Stability and Eccentricity for Two Planets in a 1:1 Resonance, and
  Their Possible Occurrence in Extrasolar Planetary Systems}.
\newblock {\em Astron. J.}, 124:2332--2338.

\bibitem[Niederman et~al., 2020]{NiedermanPousseRobutel2020}
Niederman, L., Pousse, A., and Robutel, P. (2020).
\newblock On the {{Co}}-orbital {{Motion}} in the {{Three}}-{{Body Problem}}:
  {{Existence}} of {{Quasi}}-periodic {{Horseshoe}}-{{Shaped Orbits}}.
\newblock {\em Communications in Mathematical Physics}, 377:551--612.

\bibitem[{Pierens} and {Raymond}, 2014]{PiRya2014}
{Pierens}, A. and {Raymond}, S.~N. (2014).
\newblock {Disruption of co-orbital (1:1) planetary resonances during
  gas-driven orbital migration}.
\newblock {\em MNRAS}, 442:2296--2303.

\bibitem[Pousse et~al., 2017]{PousseRobutelVienne2017}
Pousse, A., Robutel, P., and Vienne, A. (2017).
\newblock On the co-orbital motion in the planar restricted three-body problem:
  The quasi-satellite motion revisited.
\newblock {\em Celestial Mechanics and Dynamical Astronomy}, 128:383--407.

\bibitem[{Roberts}, 2002]{Robe2002}
{Roberts}, G. (2002).
\newblock Linear stability of the elliptic {L}agrangian triangle solutions in
  thethree-body problem.
\newblock {\em Journal of Dynamics and Differential Equations}, 182:191--218.

\bibitem[{Robutel} et~al., 2016]{RoNiPo2016}
{Robutel}, P., {Niederman}, L., and {Pousse}, A. (2016).
\newblock Rigorous treatment of the averaging process for co-orbital motions in
  the planetary problem.
\newblock {\em Computational and Applied Mathematics}, 35(3):951--985.

\bibitem[Robutel and Pousse, 2013]{RobutelPousse2013}
Robutel, P. and Pousse, A. (2013).
\newblock On the co-orbital motion of two planets in quasi-circular orbits.
\newblock {\em Celestial Mechanics and Dynamical Astronomy}, 117:17--40.

\bibitem[Rodr{\'i}guez et~al., 2013]{RodriguezGiupponeMichtchenko2013}
Rodr{\'i}guez, A., Giuppone, C.~A., and Michtchenko, T.~A. (2013).
\newblock Tidal evolution of close-in exoplanets in co-orbital configurations.
\newblock {\em Celestial Mechanics and Dynamical Astronomy}, 117:59--74.

\bibitem[Singer, 1968]{Singer1968}
Singer, S.~F. (1968).
\newblock The {{Origin}} of the {{Moon}} and {{Geophysical Consequences}}*.
\newblock {\em Geophysical Journal of the Royal Astronomical Society},
  15(1-2):205--226.

\bibitem[{Vokrouhlick{\'y}} and {Nesvorn{\'y}}, 2014]{VoNe2014}
{Vokrouhlick{\'y}}, D. and {Nesvorn{\'y}}, D. (2014).
\newblock {Transit Timing Variations for Planets Co-orbiting in the Horseshoe
  Regime}.
\newblock {\em APJ}, 791:6.

\bibitem[{Yoder}, 1995]{Yoder_1995cnt}
{Yoder}, C.~F. (1995).
\newblock {Astrometric and geodetic properties of Earth and the Solar System}.
\newblock In {\em Global Earth Physics: A Handbook of Physical Constants},
  pages 1--31. American Geophysical Union.

\end{thebibliography}

\newpage
\appendix
{\LARGE \textbf{Appendices}}
\section{Notations}\label{append_notation}
We gather here for convenience the notations used throughout this work\footnote{\fat{\qoppa} (qoppa) and \fat{\sampi} (sampi, see Eq. (\ref{spectrum})) are archaic Greek letters.}.
\begin{table}[H]
\begin{center}
\begin{tabular}{|l|l|}
	\hline
	\fat{m_0,R_0}&Mass and radius of the central body\\
	\hline
	\fat{m_1,R_1}&Mass and radius of the leading coorbital body\\
	\hline
	\fat{m_2,R_2}&Mass and radius of the trailing coorbital body\\
	\hline
	\fat{\mathcal{G},\,\varepsilon}&The gravitational constant, \fat{\,\left(m_1+m_2\right)/m_0}\\
	\hline
	\fat{m,\,\beta_j,\,\mu_j,\,\mu_0}&\fat{\sqrt{m_1m_2},\;\;m_0m_j/\left(m_0+m_j\right),\;\;\mathcal{G}\left(m_0+m_j\right),\;\;\mathcal{G}m_0}\\
	\hline
	\fat{a_j,\,e_j,\,\lambda_j,\,\varpi_j}&Semi-major axis, eccentricity, mean longitude, longitude of pericenter\\
	\hline
	\fat{\xi,\,\gamma,\,\delta,\,\Delta}&\fat{\lambda_1-\lambda_2,\;\;\left(m_1+m_2\right)/m,\;\;m_1/\left(m_1+m_2\right),\;\;\sqrt{2-2\cos\xi}}\\
	\hline
	\fat{\Lambda_j,\,x_j,\,\tilde{x}_j}&\fat{\beta_j\sqrt{\mu_ja_j},\;\;\sqrt{\Lambda_j}\left(1-\sqrt{1-e_j^2}\right)^{1/2}\exp\left(i\varpi_j\right),\;\;-i\bar{x}_j}\\
	\hline
	\fat{\eta,\,\bar{a},\,\tau,\,\dot{\phantom{x}},\,T}&Mean motion at the resonance, \fat{\mu_0^{1/3}\eta^{-2/3},\;\;\eta t,\;\;d/d\tau,\;\;2\pi/\eta}\\
	\hline
	\fat{J,\,J_2,\,X_j,\,Q_j}&See equations (\ref{change_var_3}), Quality factor of planet \fat{j}\\
	\hline
	\fat{\kappa_{2,j},\,\Delta t_j,\,z_j^{\bigstar}}&Second Love number, constant time lag, \fat{z_j\left(t-\Delta t_j\right)} for any \fat{z_j}\\
	\hline
	\fat{C_j,\alpha_j,\theta_j,\omega_j,\vartheta_j}&Moment of inertia, \fat{\;C_j/\left(m_jR_j^2\right)}, rotation angle, \fat{\;d\theta_j/dt,\;1-\omega_j/\eta}\\
	\hline
	\fat{\qoppa_j,\,q_j,\,\Delta\lambda_j,\,\Delta\theta_j}&\fat{R_j/\bar{a},\;\;\kappa_2^j\qoppa_j^5,\;\;\lambda_j-\lambda_j^{\bigstar},\;\;\theta_j-\theta_j^{\bigstar}}\\
	\hline
	\fat{\nu,\,g_1,\,g_2,\,x,\,y}&\fat{\sqrt{27\varepsilon/4},\;\;27\varepsilon/8,\;\;0,\;\;m_1/m_2,\;\;q_2\Delta t_2/\left(q_1\Delta t_1\right)}\\
	\hline
	\fat{\R_j,f_1,f_2}&\fat{1+f_j,\;\;m/\left( m_1+m_2\right)J_2+m/m_1J,\;\;m/\left(m_1+m_2\right)J_2-m/m_2J}\\
	\hline
\end{tabular}
\caption{Notations}\label{notation}
\end{center}
\end{table}

\section{Coefficients of the Hamiltonian \fat{\Ham_4} and \fat{\Ham_t^j}}\label{append_H4}
First we give the coefficients depending on \fat{\xi} of the Hamiltonian \fat{\Ham_4}. We recall that \fat{\Ham_4} is expressed
\begin{equation}
\begin{split}
&\Ham_4=\frac{1}{4}\frac{m}{m_0}\left\lbrace\accolade D_h\left(X_1^2\bar{X}_1^2+X_2^2\bar{X}_2^2\right)+E_hX_1^2\bar{X}_2^2+\bar{E}_hX_2^2\bar{X}_1^2\right .\\ 
& \left . +F_h\left( X_1X_2\bar{X}_1^2+\bar{X}_1\bar{X}_2X_2^2\right) +\bar{F}_h\left(\bar{X}_1\bar{X}_2X_1^2+X_1X_2\bar{X}_2^2\right) +G_hX_1X_2\bar{X}_1\bar{X}_2\accolade\right\rbrace.
\end{split}
\end{equation}
The coefficients read \fat{\left(\text{recall that }\Delta=\sqrt{2-2\cos\xi}\right)}
\begin{equation}
\begin{split}
&D_h=\frac{7}{16}\cos\xi+\frac{1}{4\Delta^9}\left(-\frac{3951}{32}+115\cos\xi+\frac{293}{8}\cos 2\xi-27\cos 3\xi-\frac{37}{32}\cos 4\xi\right), \\
&G_h=\cos\xi+\Delta^{-9}\left(-\frac{4491}{32}+139\cos\xi+\frac{233}{8}\cos 2\xi-27\cos 3\xi-\frac{25}{32}\cos 4\xi\right),\\
&E_h=\frac{1}{32}\left(e^{-i\xi}+81e^{-3i\xi}\right)+\frac{e^{-6i\xi}}{32\Delta^9}P_E\left(e^{i\xi}\right),\\
&F_h=-\frac{7}{4}e^{2i\xi}+\frac{e^{-3i\xi}}{4\Delta^9}P_F\left(e^{i\xi}\right),\\
&P_E\!\left(X\right)\!=\!-\frac{9}{8}\!+\!15X\!-\!\frac{349}{2}X^2\!+\!171X^3\!+\!\frac{2889}{4}X^4\!-\!1571X^5\!+\!\frac{2007}{2}X^6\!-\!87X^7\!-\!\frac{625}{8}X^8,\\
&P_F\!\left(X\right)\!=\!\frac{207}{32}\!+\!\frac{303}{8}X\!-\!\frac{577}{4}X^2\!+\!\frac{603}{8}X^3\!+\!\frac{2511}{16}X^4\!-\!\frac{1475}{8}X^5\!+\!45X^6\!+\!\frac{57}{8}X^7\!-\!\frac{5}{32}X^8.
\end{split}
\end{equation}
We now give the expressions of the coefficients appearing in the tidal Hamiltonian (\ref{Ht_rectangular}) for the second and fourth order in eccentricity. We have
\begin{equation}
\begin{split}
&B_t^j=\frac{3}{8}-\frac{15}{8}\cos 2\left( \Delta\lambda_j-\Delta\theta_j \right),\\
&C_t^j=\frac{3}{32}e^{i\left( \Delta\lambda_j-2\Delta\theta_j \right)}+\frac{9}{16}e^{-i\Delta\lambda_j}+\frac{147}{32}e^{-i\left( 3\Delta\lambda_j-2\Delta\theta_j \right) },\\
&D_t^j=\frac{3}{8}+\frac{69}{64}\cos 2\left( \Delta\lambda_j-\Delta\theta_j\right),\\ &G_t^j=\frac{9}{16}+\frac{75}{16}\cos 2\left( \Delta\lambda_j-\Delta\theta_j\right),\\
&E_t^j=\frac{81}{64}e^{-2i\Delta\lambda_j}+\frac{867}{32}e^{-2i\left(2\Delta\lambda_j-\Delta\theta_j\right) },\\
&F_t^j=\frac{9}{16}e^{i\Delta\lambda_j}-\frac{3}{128}e^{-i\left(\Delta\lambda_j-2\Delta\theta_j\right) }-\frac{1365}{128}e^{i\left(3\Delta\lambda_j-2\Delta\theta_j\right) },
\end{split}
\end{equation}
with
\begin{equation}
\Delta\lambda_j=\lambda_j-\lambda_j^{\bigstar}\;\;\text{ and }\;\;\Delta\theta_j=\theta_j-\theta_j^{\bigstar}.
\end{equation} 

\section{Lagrange and anti-Lagrange in horseshoe-shaped orbits}\label{append_hs}
Here we show that the Lagrange and anti-Lagrange proper modes correspond respectively to aligned and anti-aligned pericenters in horseshoe-shaped orbits. The matrix of the variational equations (\ref{eq:limX}), once averaged over the semi fast dynamics and according to the geometrical considerations stated at the end of section \ref{conservative_excentrique}, reads
\begin{equation}
\mathcal{M}_0=-i\begin{pmatrix}\frac{m_2}{m_0}\underline{A}_h&\frac{m_2}{m_0}\underline{B}_h\vspace{1mm}\\
\frac{m_1}{m_0}\underline{B}_h&\frac{m_1}{m_0}\underline{A}_h
\end{pmatrix},
\end{equation}
where both \fat{\underline{A}_h} and \fat{\underline{B}_h}, average of \fat{A_h} and \fat{B_h} over the semi-fast dynamics, are real. The eigenvectors of \fat{\mathcal{M}_0} show that for the Lagrange configuration
\begin{equation}\label{zero}
\arg\left(\frac{X_1}{X_2}\right)=\varpi_1-\varpi_2=0,
\end{equation}
while for the anti-Lagrange configuration
\begin{equation}\label{pi}
\arg\left(\frac{X_1}{X_2}\right)=\varpi_1-\varpi_2=\pi.
\end{equation}
This corresponds to aligned and anti-aligned pericenters.

\section{Conservation of the total angular momentum}\label{append_angular}

Here we show that the set of equations (\ref{equation_case1}) is consistent with the conservation of the total angular momentum of the system.
The normalized \footnote{Normalized by \fat{m\bar{a}^2\eta}.} total angular momentum \fat{\mathcal{C}} reads \citep{RobutelPousse2013}
\begin{equation}
\mathcal{C}=\sum_j\frac{m_j}{m}\R_j-\frac{1}{2}\sum_j\frac{m_j}{m}X_j\bar{X}_j+\sum_j\alpha_j\frac{m_j}{m}\qoppa_j^2\left(1-\vartheta_j \right).
\end{equation}
From (\ref{equation_case1}) we get
\begin{equation}
\dot{\mathcal{C}}=\sum_{j\in\left\lbrace 1,2\right\rbrace}3\frac{q_j}{Q_j}\frac{m_0}{m}\R_j^{-13}X_j\bar{X}_j\left\lbrace\accolade h_2^j-k_2^j+p_2^j+\R_j^{-1}X_j\bar{X}_j\left(h_4^j-k_4^j+p_4^j \right)\accolade\right\rbrace,
\end{equation}
and the total angular momentum is conserved since we have
\begin{equation}
\begin{split}
&h_2^j-k_2^j+p_2^j=0\text{ and}\\
&h_4^j-k_4^j+p_4^j=0.
\end{split}
\end{equation}

\section{Diagonalization of a perturbed matrix}\label{append_diago}
We show here the method that we use to obtain the eigenvalues of a perturbed matrix once a diagonal basis of the principal matrix is known. Indeed, the computation of the eigenvalues (\ref{eigenvalues_circular}) is equivalent to finding the roots of the characteristic polynomial of \fat{\mathcal{Z}_0+\mathcal{Z}_1}, given in equations (\ref{Z0}) and (\ref{Z1}). Even when the degree of this polynomial is reduced to four using the \fat{0} eigenvalue, it is hard to obtain its roots in a convenient form. The method we use here, briefly presented by \citep{LaskarBoueCorreia2012}, gives the eigenvalues and eigenvectors very easily.\\\\
Let \fat{\mathcal{M}=\mathcal{M}_0+\zeta\mathcal{M}_1\in\mathcal{M}_n\left(\mathbb{C}\right)} be a \fat{n\times n} complex matrix where \fat{\zeta} is a small quantity with respect to \fat{1}. Assume that we know a diagonal basis for \fat{\mathcal{M}_0}
\begin{equation}
\mathcal{D}_0=P_0^{-1}\mathcal{M}_0P_0=\text{diag}\left(\lambda_i\right),
\end{equation}
where the columns of \fat{P_0} are the eigenvectors of \fat{\mathcal{M}_0} and the \fat{\lambda_i} its eigenvalues, which are not assumed to be of multiplicity one but which are assumed to be sorted by value, that is, equal eigenvalues are consecutive. This does not restrict the generality, as any permutation can be applied on the columns of \fat{P_0} to achieve that. We now define 
\begin{equation}
\mathcal{Q}_1=P_0^{-1}\mathcal{M}_1P_0.
\end{equation}
If \fat{P} is the matrix of the eigenvectors of \fat{\mathcal{D}_0+\zeta\mathcal{Q}_1}, and since \fat{\mathcal{D}_0+\zeta\mathcal{Q}_1} is near diagonal, we write
\begin{equation}
P=I_n+\zeta P_1+\mathcal{O}\left(\zeta^2\right).
\end{equation}
We have
\begin{equation}
P^{-1}\left(\mathcal{D}_0+\zeta\mathcal{Q}_1\right)P=\mathcal{D}_0+\zeta\left(\accolade\mathcal{Q}_1+\left[\mathcal{D}_0,P_1\right]\accolade\right)+\mathcal{O}\left(\zeta^2\right),
\end{equation}
where \fat{\left[\mathcal{D}_0,P_1\right]=\mathcal{D}_0P_1-P_1\mathcal{D}_0}. Thus the cohomological equation
\begin{equation}
\mathcal{Q}_1+\left[\mathcal{D}_0,P_1\right]=\mathcal{D}_1,
\end{equation}
where \fat{\mathcal{D}_1=\text{diag}\left(q_{i,i}\right)} is the diagonal matrix composed of the diagonal terms of \fat{\mathcal{Q}_1}. The solution of the cohomological equation is
\begin{equation}
p_{i,j}=\left\lbrace\begin{matrix}\frac{q_{i,j}}{\lambda_j-\lambda_i}&\text{ if }\lambda_i\neq\lambda_j,\vspace{1mm}\\0&\text{ else},\end{matrix}\right.
\end{equation}
where
\begin{equation}
\mathcal{Q}_1=\left(q_{i,j}\right)_{1\leq i,j\leq n}\place\text{and}\place P_1=\left(p_{i,j}\right)_{1\leq i,j\leq n}.
\end{equation}
The matrix \fat{\mathcal{M}_0+\zeta\mathcal{M}_1} is now block diagonal :
\begin{equation}
P^{-1}P_0^{-1}\left(\mathcal{M}_0+\zeta\mathcal{M}_1\right)P_0P=\text{diag}\left(\mathcal{B}_0^1+\zeta\mathcal{B}_1^1,\,...\,,\mathcal{B}_0^r+\zeta\mathcal{B}_1^r\right),\place r\leq n
\end{equation}
where \fat{\forall i\leq r\;\;\exists k\leq n} such that
\begin{equation}
\mathcal{B}_0^i=\lambda_kI_{m(k)},
\end{equation}
and \fat{m(k)} denotes the multiplicity of \fat{\lambda_k} and thus the size of the block. The computation of the eigenvalues of \fat{\mathcal{M}_0+\zeta\mathcal{M}_1} is reduced to the computation of the eigenvalues of the blocks \fat{\mathcal{B}_0^i+\zeta\mathcal{B}_1^i} who are hopefully all of small size and whose eigenvalues are then analytically easily found.

\section{Linearization}\label{append_linear}
Near the fixed points given by (\ref{point_fixe_case1}), the linear system reads
\begin{equation}
\dot{\mathcal{X}}=\left(\mathcal{Q}_0+\mathcal{Q}_1\right)\mathcal{X}.
\end{equation}
Where \fat{\mathcal{X}} is defined in section \ref{linearization}. We have
\begin{equation}
\mathcal{Q}_0=\begin{pmatrix}\mathcal{Z}_0&0_{5,2}\\0_{2,5}&\mathcal{M}_0\end{pmatrix}\;\;\text{ and }\;\;\mathcal{Q}_1=\begin{pmatrix}\mathcal{Z}_1&0_{5,2}\\0_{2,5}&\mathcal{M}_1\end{pmatrix},
\end{equation}
where
\begin{equation}\label{M0}
\mathcal{M}_0=\frac{27}{8}i\begin{pmatrix}\frac{m_2}{m_0}&-\frac{m_2}{m_0}e^{i\pi/3}\vspace{1mm}\\-\frac{m_1}{m_0}e^{-i\pi/3}&\frac{m_1}{m_0}\end{pmatrix},
\end{equation}
\begin{equation}\label{M1}
\mathcal{M}_1=-\frac{21}{2}\text{diag}\left\lbrace\accolade q_1\frac{m_0}{m_1}\left(\eta\Delta t_1-\frac{5}{7}i\right),q_2\frac{m_0}{m_2}\left(\eta\Delta t_2-\frac{5}{7}i\right)\accolade\right\rbrace,
\end{equation}
\begin{equation}\label{Z0}
\mathcal{Z}_0=\begin{pmatrix}
0&0&0&0&0\\
0&0&0&0&0\\
0&0&0&-3\gamma&0\\
0&0&\frac{1}{3}\gamma^{-1}\nu^2&0&0\\
0&0&0&0&0
\end{pmatrix},
\end{equation}
\begin{equation}\label{Z1}
\mathcal{Z}_1=\begin{pmatrix}
-d_1 & 0 & 0 & 3\gamma^{-1}\delta^{-1}d_1 & 3\gamma^{-1}d_1 \\
0 & -d_2 & 0 & -3\gamma\delta d_2 & 3\gamma^{-1}d_2 \\
0 & 0 & 0 & -\gamma\left(\delta c_2+\left(1-\delta\right)c_1\right)  & \gamma^{-1}\left(c_2-c_1\right)  \\
-\left(1-\delta \right)b_1 & \delta b_2 & 0 & 3\gamma\left(\delta^2b_2+\left(1-\delta \right)^2b_1  \right) & 3\gamma^{-1}\left[ \left(1-\delta \right)b_1-\delta b_2\right]  \\
-b_1 & -b_2 & 0 & 3\gamma^{-1}\delta^{-1}b_1-3\gamma\delta b_2 & 3\gamma^{-1}\left(b_1+b_2 \right) 
\end{pmatrix},
\end{equation}
with
\begin{equation}
\begin{split}
&\delta=\frac{m_1}{m_1+m_2},\;\;\;\;\;\;\gamma=\frac{m_1+m_2}{m},\;\;\;\;\;\;\nu=\sqrt{\frac{27\varepsilon}{4}},\\
&d_j=\frac{3}{\alpha_j}\frac{q_j}{Q_j}\qoppa_j^{-2}\frac{m_0}{m_j},\place b_j=3\frac{q_j}{Q_j}\frac{m_0}{m},\place c_j=78q_j\frac{m_0}{m_j}.
\end{split}
\end{equation}
Near \fat{L_{4,5}}, the eigenvectors of \fat{\mathcal{M}_0+\mathcal{M}_1}, computed using results from appendix \ref{append_diago} reveal that the Lagrange configuration corresponds to
\begin{equation}
\begin{split}
&\varpi_1-\varpi_2=\frac{\pi}{3}+\frac{28}{9}\frac{m_0^2\left(m_1q_2/Q_2+m_2q_1/Q_1\right)}{m_1m_2\left(m_1+m_2\right)},\\&\frac{e_1}{e_2}=1+\frac{20}{9}\frac{m_0^2\left(q_2m_1-q_1m_2\right)}{m_1m_2\left(m_1+m_2\right)},
\end{split}
\end{equation}
while the anti-Lagrange configuration complies with
\begin{equation}
\begin{split}
&\varpi_1-\varpi_2=\frac{4\pi}{3}-\frac{28}{9}\frac{m_0^2\left(m_1q_2/Q_2+m_2q_1/Q_1\right)}{m_1m_2\left(m_1+m_2\right)},\\&\frac{e_1}{e_2}=\frac{m_2}{m_1}\left(1-\frac{20}{9}\frac{m_0^2\left(q_2m_1-q_1m_2\right)}{m_1m_2\left(m_1+m_2\right)}\right).
\end{split}
\end{equation}

\section{Direct $3$-body model}\label{append_nbody_direct}

The complete equations of motion governing the tidal evolution of a three-body system in an astrocentric frame using a linear constant time-lag tidal model  are given by \citep{Mignard1979}

\begin{equation}\label{nbody_direct}
\begin{split}
& \frac{d^2{\vec{r}}_1}{dt^2}= -\frac{\mu_1}{r_1^3}\vec{r}_1+ \mathcal{G} m_2\left(\frac{\vec{r}_2-\vec{r}_1}{|\vec{r}_2-\vec{r}_1|^3}-\frac{\vec{r}_2}{r_2^3}\right) + \frac{\vec{f}_1}{\beta_1} + \frac{\vec{f}_2}{m_0}  \ , \\
& \frac{d^2{\vec{r}}_2}{dt^2} = -\frac{\mu_2}{r_2^3}\vec{r}_2+ \mathcal{G} m_1\left(\frac{\vec{r}_1-\vec{r}_2}{|\vec{r}_1-\vec{r}_2|^3}-\frac{\vec{r}_1}{r_1^3}\right) + \frac{\vec{f}_2}{\beta_2} + \frac{\vec{f}_1}{m_0} \ , \\
& \frac{d^2{\theta}_i}{dt^2} = - \frac{(\vec{r}_i \times \vec{f}_i)\cdot \vec{k}}{C_i} = - 3 \frac{\kappa_{2, i} {\cal G} m_0^2 R_i^3}{\alpha_i m_i r_i^{8}} \Delta t_i \left[  \frac{d{\theta}_i}{dt} \, r_i^2   - \left( \vec{r}_i \times\frac{d{\vec{r}}_i}{dt} \right) \cdot \vec{k} \right]   \ ,
\end{split}
\end{equation}
where $\vec{r}_i$ and $\theta_i$ are the astrocentric position vector and the rotation angle of the planet $i$, respectively,
$\vec{k}$ is the unit vector normal to the orbital plane of the planets, and $\vec{f}_i$ is the force arising from the tidal potential energy created by the deformation of each planet (Eq.\,(\ref{perturbed_energy}))
\begin{equation}\label{fdef}
\vec{f}_i = - 3 \frac{\kappa_{2, i} {\cal G} m_0^2 R_i^5}{r_i^{8}} \vec{r}_i -3 \frac{\kappa_{2, i} {\cal G} m_0^2 R_i^5}{r_i^{10}} \Delta t_i \left[ 2 \left( \vec{r}_i \cdot \frac{d{\vec{r}}_i}{dt} \right) \vec{r}_i + r_i^2 \left( \frac{d{\theta}_i}{dt} \, \vec{r}_i \times \vec{k} + \frac{d{\vec{r}}_i}{dt} \right)\right] \ .
\end{equation}

\end{document}